\documentclass[aps,prb,floatfix,amsmath,amssymb,preprint,eqsecnum]{revtex4}
\usepackage{graphicx}
\usepackage{dcolumn}
\usepackage{bm}
\usepackage{epstopdf}
\usepackage{setspace}   
\usepackage{amsmath}
\usepackage{amssymb}

\preprint{arXiv:0901.0005}

\begin{document}
\title{Paired electron pockets in the hole-doped cuprates}

\author{Victor Galitski}
\affiliation{
Joint Quantum Institute and CNAM, Department of Physics,
University of Maryland, College Park, MD 20742-4111}

\author{Subir Sachdev}
\affiliation{Department of Physics, Harvard University, Cambridge MA
02138}

\date{December 30, 2008\\
\vspace{1.6in}}
\begin{abstract}
We propose a theory for the underdoped hole-doped cuprates,
focusing on the ``nodal-anti-nodal dichotomy'' observed in recent
experiments. Our theory begins with an ordered antiferromagnetic
Fermi liquid with electron and hole pockets. We argue that it is useful
to consider a quantum transition at which the loss of
antiferromagnetic order leads to a hypothetical metallic ``algebraic charge liquid'' (ACL)
with pockets of charge $-e$ and $+e$ fermions, and an emergent
U(1) gauge field; the instabilities of the ACL lead to the low temperature
phases of the underdoped cuprates.
The pairing instability leads to a superconductor with the strongest pairing within
the $-e$ Fermi pockets, a $d$-wave pairing signature for electrons, and very
weak nodal-point pairing of the $+e$ fermions near the Brillouin
zone diagonals. The influence of an applied magnetic field is discussed
using a proposed phase diagram as a function of field strength and doping. We
describe the influence of gauge field and pairing
fluctuations on the quantum Shubnikov-de~Haas oscillations in the
normal states induced by the field. For the finite temperature pseudogap region,
our theory has some similarities to the phenomenological two-fluid
model of $-2e$ bosons and $+e$ fermions proposed by Geshkenbein, Ioffe, and
Larkin [Phys. Rev. B {\bf 55}, 3173 (1997)], which describes anomalous aspects of
transverse transport in a magnetic field.
\end{abstract}

\maketitle

\section{Introduction}
\label{sec:intro}

\begin{figure}
\includegraphics[width=4.0in]{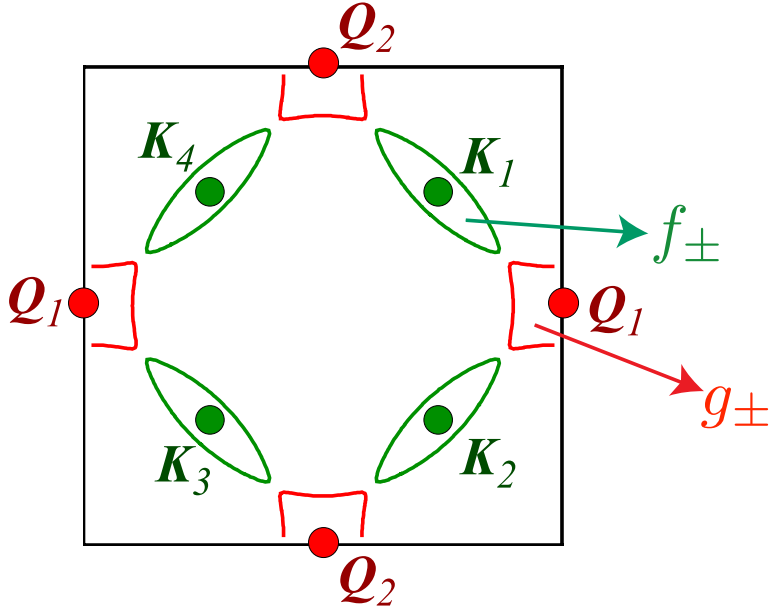}
\caption{Square lattice Brillouin zone map showing the electron and hole pockets
of conventional spin density wave theory \cite{sokol,morr,millisnorman,harrison} of antiferromagnetic order at
wavevector $(\pi,\pi)$.  The hole pockets are
centered at the  wavevectors
${\bf K}_v = (\pm \pi/2, \pm \pi/2)$ (where $v=1\ldots 4$), and
the electron pockets are centered at ${\bf
Q}_a = (\pi,0),(0,\pi)$ (where $a=1,2$).
The present paper describes the influence of quantum and thermal fluctuations in the orientation
of the antiferromagnetic order on the electron and hole pockets. We will find strong pairing of the
electron pockets at ${\bf Q}_a$, which induces a weak `proximity effect' pairing of the hole pockets
at ${\bf K}_v$, with all the pairings consistent with a $d_{x^2-y^2}$ pairing signature.}
\label{fig:bz}
\end{figure}

A remarkable consensus has emerged in recent experiments
\cite{shen1,kanigel,letacon,shen2,kohsaka1,hudson,kohsaka2,vidya,sawatzky}
on the enigmatic underdoped region of the hole-doped cuprate
superconductors. These experiments reveal a clear ``dichotomy''
between the low-lying electronic excitations near the nodal points
of the $d$-wave superconductor ({\em i.e.\/} near the wavevectors
${\bf K}_v$ in Fig.~\ref{fig:bz}) and the higher energy
excitations near the ``anti-nodal'' points ({\em i.e.\/} near the
wavevectors ${\bf Q}_a$ in Fig.~\ref{fig:bz})
The nodal quasiparticles have a pairing energy which decreases with decreasing doping, and they form coherent quasiparticles
which display characteristic interference patterns in scanning tunneling microscopy (STM) observations.
In contrast, the anti-nodal excitations have a larger gap which increases with decreasing doping,
and they appear to be excitations of a state with STM modulations characteristic of a valence bond solid.~\cite{kohsaka1}

Theoretically, a number of numerical studies
\cite{honerkamp,senechal,civelli} of the Hubbard model have also
presented evidence for the nodal-anti-nodal dichotomy at
intermediate energy scales. These results are connected to ad hoc
theoretical models  \cite{geshkenbein,konik,yrz,tsvelik} involving ``Fermi
arc'' and/or electron/hole pockets which violate the traditional
Luttinger theorem on the area enclosed by the Fermi surfaces. A
central point behind the analysis of the present paper is that
theories with such ad hoc violations of the Luttinger Fermi area
law are fundamentally incomplete. Using arguments building upon
the non-perturbative proof of the Luttinger theorem,
\cite{oshikawa} it was argued \cite{senthilvojta1,senthilvojta2}
that metallic states with non-Luttinger Fermi surfaces must have
``topological order,'' by which we mean there must be additional
collective excitations associated with an emergent gauge field.
Such collective excitations are crucial in the
description of such exotic conducting states. Specific theories
\cite{rkk1,rkk2,rkk3} of conducting states with non-Luttinger
areas, labeled ``algebraic charge liquids'' (ACL), have been
provided: in models appropriate for the cuprates, these states
were obtained across quantum transitions involving the loss of
antiferromagnetic  N\'eel order. Furthermore, the formalism
developed to describe the quantum ACL is also useful
for describing the ``liquid'' state obtained when the antiferromagnetic order
is lost by thermal fluctuations.

We begin presentation of our results by recalling
spin-density-wave (SDW) studies of the onset of antiferromagnetic order in the doped
cuprates.~\cite{sokol,morr} These works are expressed in terms of a
vector SDW order parameter $N_\ell$ (with $\ell = x, y, z$),
measuring the spin-density-wave at wavevector $(\pi,\pi)$, which
can perturb the Fermi surface of a weak-coupling band structure; we will
restrict our attention here to the commensurate N\'eel SDW, and
recent work by Harrison \cite{harrison} has shown that a similar Fermi surface
structure is obtained for incommensurate SDW order.
The theory for the transition from the SDW ordered state ($\langle
N_\ell \rangle \neq 0$) to the non-magnetic state ($\langle N_\ell
\rangle = 0$) is expressed in terms of an effective action for
spacetime fluctuations $N_\ell$. The state with $\langle N_{\ell}
\rangle \neq 0$ has ``small'' Fermi pockets: hole pockets centered
at the ${\bf K}_v$ and electron pockets centered at the ${\bf
Q}_a$. This conventional, ordered antiferromagnetic state will
also be present in our theory below. In the spin-density-wave
theory, the non-magnetic state with $\langle N_\ell \rangle =0$
has a ``large'' Fermi surface which obeys the conventional
Luttinger theory, and the transition from
the small Fermi pockets state to a ``large'' Fermi surface {\em
co-incides\/} with the loss of SDW order.

\begin{figure}
\includegraphics[width=4.5in]{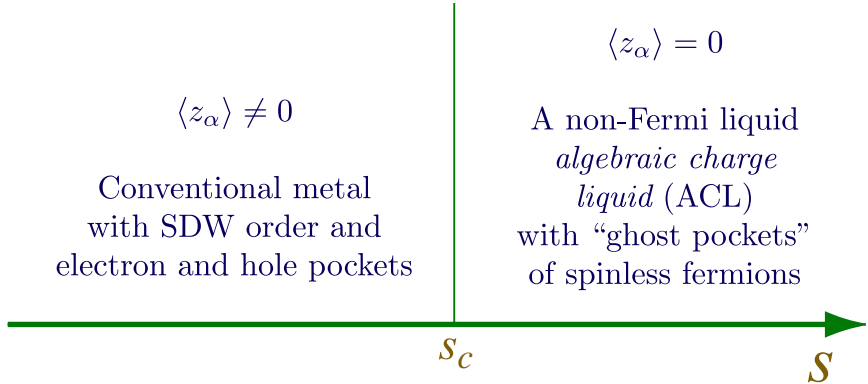}
\caption{Non-superconducting ground states of our theory as a function of the coupling $s$ which tunes the
strength of the SDW fluctuations (see $\mathcal{L}_z$
in Eq.~(\ref{Lz})); $s$ is controlled by varying the doping $\delta$ and by choosing different cuprate series.
We expect that at zero applied magnetic field, $H$, these phases are pre-empted by superconductivity;
the phase diagram as a function of $H$ and $\delta$ is shown in Fig.~\ref{fig:aclpd2}.
}
\label{fig:aclpd1}
\end{figure}

In our theory below, the physical properties of the SDW ordered
state are qualitatively identical to those in the
spin-density-wave theory. However, we express our theory for the
loss of SDW order not in terms of the vector $N_\ell$ order
parameter, but in terms of a bosonic spinor $z_\alpha$ which is
related to $N_\ell$ by
\begin{equation}
N_\ell = z^\ast_{\alpha}
\sigma^\ell_{\alpha\beta} z_\beta,
\end{equation}
where the $\sigma^\ell$ are
the Pauli matrices. Then, the  state with $\langle z_\alpha
\rangle \neq 0$ is the same as the spin-density wave state with
$\langle N_\ell \rangle \neq 0$. However, an important advantage of the formulation
in terms of the $z_\alpha$ is that we can describe the electron spin in
terms of its components quantized along the direction of the local N\'eel order,
simply by performing a SU(2) rotation defined by the spinor $z_\alpha$.
This facilitates a description of the non-magnetic state
\cite{rkk1,rkk2,rkk3} with $\langle z_\alpha \rangle = 0$, which is a
topologically ordered ACL that retains key aspects of the
``small'' Fermi surface structure, as summarized in Fig.~\ref{fig:aclpd1}, and will be discussed
in detail below.
The $z_\alpha$ formalism also efficiently describes the non-magnetic state
obtained when SU(2) invariance is restored by thermal fluctuations.

\begin{figure}
\includegraphics[width=4.5in]{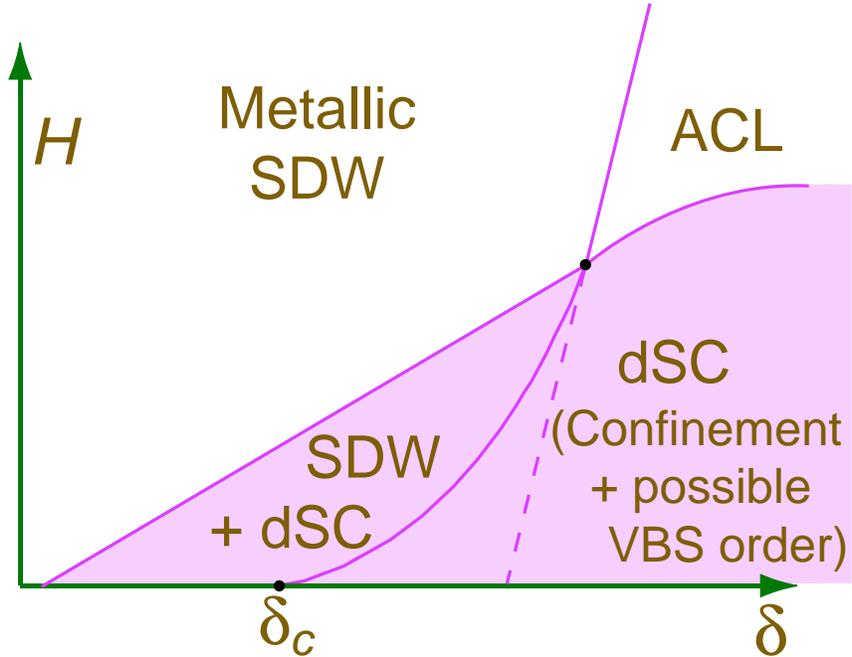}
\caption{Proposed zero temperature phase diagram for the underdoped cuprates as a function of doping, $\delta$,
and a magnetic field $H$
perpendicular to the layers; the insulating phases at very small $\delta$ are not shown.
This phase diagram combines our present results with those of Refs.~\onlinecite{sokol} and~\onlinecite{demler}. The value
of $\delta_c$, and the scale of $\delta$ will be differ for the various cuprate series.
The superconducting phases are shaded and labeled dSC. The dashed line indicates the normal state phase boundary
of Fig.~\ref{fig:aclpd1} which is pre-empted by the onset of superconductivity.
The physics of the confinement and possible valence bond solid (VBS) order in the dSC state is discussed in Ref.~\onlinecite{rkk3}.
At higher $\delta$, there is a transformation to
the physics of the ``large'' Fermi surface state, which we do not describe here.
Only the ACL phase above has unconventional `topological' order; all other phases have conventional order
and associated excitations.
While the ACL can be a stable ground state at large $H$, it is possible it is pre-empted by
a conventional Fermi liquid. }
\label{fig:aclpd2}
\end{figure}

The primary motivation for the present paper comes from the recent
experimental evidence for electron pockets in
the hole doped cuprates at large magnetic fields: Shubnikov-de Hass (SdH) and de Hass-van Alphen (dHvA) oscillations \cite{doiron,cooper,nigel,cyril,suchitra}
indicate carriers in small pockets, and Hall conductivity measurements \cite{louis} have been used to argue that
these are electron pockets.
One of our main claims is that an algebraic charge liquid consisting of a pocket of charge $-e$ fermions at the ${\bf Q}_a$ wavevectors,
and of charge $+e$ fermions at the ${\bf K}_v$ wavevectors, provides the underlying quantum state for the description
of the underdoped cuprates, and also for the thermal fluctuations of the more classical liquid state in the `pseudogap' regime.
Speculations along these lines were also made in Ref.~\onlinecite{rkk3}.
This ACL preserves the full symmetry of the Hamiltonian. Instabilities
of this ACL involving the onset of SDW order, superconductivity, and charge order will be key for the description
of the underdoped regime --- see the proposed phase diagram in Fig.~\ref{fig:aclpd2}.
 In particular, the onset of superconductivity removes low energy fermionic excitations which
suppress monopole-instantons in the gauge field, and so is likely to lead to a confinement transition; this confinement physics
has been studied in model systems in earlier work \cite{rkk3}, and we will not discuss it further here.

We shall pay
particular attention here to the pairing of the $-e$ pocket. We note that a phenomenological
model of pairs of electrons near the ${\bf Q}_a$ wavevectors, with charge $-2e$,
was considered by Geshkenbein, Ioffe, and Larkin,~\cite{geshkenbein}
and we will discuss the connection to their model further in Section~\ref{sec:model}.
As we will describe there, the  charge carriers in this pocket experience a strong attractive interaction,
which causes them to form an $s$-wave paired state. However,
the resulting superconducting state actually has $d$-wave pairing for the physical electrons,~\cite{rkk3} as will be reviewed in Section~\ref{sec:model}.
Furthermore, we will show that the pairing of the fermions near the ${\bf K}_v$ wavevectors is very weak, and has
nodal points along the Brillouin zone diagonals. Thus our theory contains a low density
of strongly paired charge carriers, a $d$-wave pairing signature with nodal points, and a nodal-anti-nodal dichotomy:
these are all attractive phenomenological features.

We shall describe the loss of superconductivity in this state by an applied magnetic field, $H$, and the appearance of
SdH oscillations at $H > H_{c2}$, using the phase diagram in Fig.~\ref{fig:aclpd2}. Both the SDW metal and the ACL metal exhibit SdH oscillations.
Specific predictions for the $H$ and temperatures ($T$) dependence of these
SdH oscillations will be provided, which can potentially be compared with experiments.
Because our charge $\pm e$ fermions do not carry spin, a key feature of the SdH oscillations will be the absence
of a linear Zeeman splitting in the Fermi surface areas, both in the ACL and in the phase with SDW order.\cite{zeeman}

Another effect of the applied $H$ field
is that it can induce or enhance the SDW order for the fields at which superconductivity is present --- see Fig.~\ref{fig:aclpd2}. This has been discussed
theoretically in some detail,~\cite{demler} and in Section~\ref{sec:model} we will show that our present
pairing theory provides a specific mechanism for the competition between antiferromagnetism and superconductivity.
This field-induced enhancement of SDW order is
also clearly observed in the La series of hole-doped cuprates.~\cite{katano,lake,jtran2,boris,mesot,mesot2}
The samples with $\delta = 0.1$ display \cite{lake} the behavior expected for $\delta < \delta_c$;
the samples with $\delta = 0.144$ have observed \cite{boris,mesot,mesot2} the physics predicted \cite{demler} for $\delta > \delta_c$,
including the dSC to dSC+SDW transition at a non-zero $H$. Thus for the La compounds we expect that $\delta_c$ is between
the quoted values. Very recently, field induced SDW order characteristic of $\delta > \delta_c$ has also been reported \cite{mesot3} for YBa$_2$Cu$_3$O$_{6.45}$. The quantum oscillation
experiments \cite{doiron,louis,cooper,nigel,cyril,suchitra} are on the same (or closely related) compound,
and so it is plausible that they are in the metallic SDW phase of Fig.~\ref{fig:aclpd2}, as others have also suggested.\cite{fuchun}
Our predictions for quantum oscillations and pairing instabilities
will also extend to this metallic SDW phase.

Our theory also offers a natural starting point for a
description of the finite $T$ ``pseudogap'' region of the
underdoped cuprates. The strongly paired $-e$ pockets lead to an
effective description in terms of charge $-2e$ bosons which can
exist above the superconducting critical temperatures.~\cite{geshkenbein}
After a duality mapping, this leads to a theory
of a vortex liquid which can capture both the ``phase''
fluctuations and the possible instabilities to varieties of charge
order.~\cite{bbbss,markus} We will explore some of the thermoelectric transport properties
of such a model in Section~\ref{sec:propose}.

The physics of the lightly-doped Mott insulator as described by a $t$-$J$ model is usually implicated in the description
of cuprate superconductivity.~\cite{pwa} From this perspective, our use, following a proposal in Ref.~\onlinecite{rkk3},
of charge $-e$ carriers for the hole-doped cuprates
may seem unacceptable. It is often argued that such
excitations are only present across a Hubbard gap of energy $U$, and the limit $U \rightarrow \infty$ has been
taken by a Gutzwiller projection of such carriers.~\cite{pwa} In response to this potential objection, we have the following
responses:\\
({\em i\/}) We draw the reader's attention to the
electron-doped cuprates, which presumably have a similar value of $U$, and for which {\em both\/} electron-like
and hole-like carriers have been observed in photoemission experiments on the state with SDW order.~\cite{armitage03,claesson04,matsui05,park07}
We are assuming here that similar physics
applies to the hole-doped cuprates. \\
({\em ii\/}) While the upper Hubbard band is indeed separated by
an energy of order $U$, a ``Kondo resonance'' of these states can
be present at the Fermi level, and this is described by our charge
$-e$ carriers. Indeed states from both the upper and lower Hubbard
bands are required \cite{qi,rkk3} for the eventual appearance of
the Fermi-liquid Luttinger (``large'') Fermi surface state in the
overdoped
regime, and so we claim it is not surprising that both bands have precursors in the underdoped regime.\\
({\em iii\/}) We note recent arguments by Comanac {\em et al.},~\cite{comanac} based upon optical
conductivity data and dynamic mean field theory arguments,
that the effective $U$ in the hole-doped cuprates is not as large as is commonly assumed.\\
({\em iv\/})  As we will review below, the electron pockets reside in a region of the Brillouin zone where the pairing
force is strongest. Thus for these momenta, it will pay to acquire states from the upper Hubbard
band to benefit from the increased pairing energy. The effective density of carriers in the charge $-e$ pockets
will be larger than would have been assumed without accounting for the pairing. Indeed, even if the chemical potential
is below the bottom of the band of $-e$ carriers, the pairing interaction will induce a non-zero density of $-e$ carriers.

The outline of the remainder of our paper is as follows. In Section~\ref{sec:model}, we describe our model for the underdoped
cuprates. The primary actors are the electron pockets near ${\bf Q}_a$ which pair strongly. We show how this
model leads naturally to a $d$-wave superconducting pairing for physical electrons,
with only weak pairing and nodal excitations along
the zone diagonals near the ${\bf K}_v$ points. Section~\ref{sec:qosc} will consider quantum oscillations
in the normal state obtained by applying a strong magnetic field. We will describe the corrections to the
Lifshitz-Kosevich
formula from gauge field and pairing fluctuations.
Section~\ref{sec:propose} will discuss additional
experimental consequences of our theory: a two-fluid model for transverse thermoelectric transport
in the pseudogap region.

\section{The Model}
\label{sec:model}

The starting point of our analysis is an expression \cite{rkk1,rkk2,rkk3} for the electron annihilation operator
$\Psi_{\alpha} ({\bf r})$ (where $\alpha = \uparrow, \downarrow$ is a spin index) in terms of continuum fermionic fields
$F_{v\alpha}$ and $G_{a \alpha}$,
which reside in the vicinity of the wavevectors ${\bf K}_v$ and ${\bf Q}_a$ respectively:
\begin{equation}
\Psi_\alpha ({\bf r}) = \sqrt{Z_f} \sum_{v=1}^4 e^{i {\bf K}_v \cdot {\bf r} } F_{v \alpha}^\dagger +
\sqrt{Z_g} \sum_{a=1}^2 e^{i {\bf Q}_a \cdot {\bf r} } G_{a \alpha} .
\label{Psi}
\end{equation}
Here $Z_{f,g}$ are non-singular quasiparticle renormalization factors which depend upon microscopic details.
As will be described explicitly below, the fermions $F_{v\alpha}$ and $G_{a \alpha}$ are in turn expressed in terms of a bosonic spinon field $z_\alpha$,
and spinless fermions which carry the electromagnetic charge.
The phases with $\langle z_\alpha \rangle \neq 0$
have been shown \cite{rkk2} to be conventional SDW-ordered states:~\cite{sokol,morr} the excitation spectrum co-incides with
that obtained in spin-wave/Hartree-Fock theory. However, the utility of the parameterizations below is that the
same formalism can be easily extended across the quantum transition at which N\'eel order is lost,
and we reach a phase with $\langle z_\alpha \rangle =  0$.
There is substantial recent numerical evidence \cite{sandvik,melkokaul,wiese,kuklov}
that the $z_\alpha$-based theory correctly captures the low energy fluctuations across the N\'eel-disordered
transition in insulating model systems. One of our main assumptions will be that this description in terms
of the $z_\alpha$ is a valid starting point for describing the loss of N\'eel order in the doped cuprates. The theory
for this N\'eel disordering transition also involves an emergent U(1) gauge field $A_\mu \equiv (A_\tau, {\bf A})$, which is connected
to the gauge field of the CP$^1$ model; the $z_\alpha$ carry unit charge under $A_\mu$. Here $\mu$ is a spacetime index
extending over the spatial co-ordinates $x$,$y$ and the imaginary time co-ordinate $\tau$.

For the electronic excitations near the ${\bf K}_v$, we need the
electromagnetic charge $+e$ ``holon'' annihilation operators
$f_{q p}$, where $q=\pm$ and $p=1,2$. Here $q$ is the ``charge''
under $A_\mu$, and $p$ is a ``valley'' index; note that although
there are 4 pockets near the ${\bf K}_v$ in Fig.~\ref{fig:bz}, a
proper counting of degrees of freedom requires only two valleys.
The complete expressions for the $F_{v \alpha}$ at all for ${\bf
K}_v$ points in terms of the $f_{\pm p}$ are \cite{rkk1}
\begin{equation}
\left( \begin{array}{c} F_{1,2\uparrow}^\dagger \\ F_{1,2\downarrow}^\dagger \end{array} \right) = \mathcal{R}_z \left( \begin{array}{c} f_{+1,2}^\dagger \\ -f_{-1,2}^\dagger \end{array} \right)~~~~~,~~~~~
\left( \begin{array}{c} F_{3,4\uparrow}^\dagger \\ F_{3,4\downarrow}^\dagger \end{array} \right) = \mathcal{R}_z \left( \begin{array}{c} f_{+1,2}^\dagger \\ f_{-1,2}^\dagger \end{array} \right),
\label{F}
\end{equation}
where
\begin{equation}
\mathcal{R}_z \equiv \left( \begin{array}{cc}
z_\uparrow & -z_\downarrow^\ast  \\
z_\downarrow &  z_\uparrow^\ast
 \end{array} \right).
 \label{Rz}
\end{equation}
The physical content of this parameterization is simple: the $\pm$ indices of the $f_{\pm p}$ are
the spin components quantized along the local SDW order, and these
are rotated by the SU(2) matrix $\mathcal{R}_z$ to a fixed quantization direction by the $z_\alpha$.
Note that pockets separated by the SDW ordering wavevector of $(\pi,\pi)$ are parameterized by the same
degrees of freedom, and they differ only sign of the $f_{-v}$ operator.
Eq.~(\ref{F}) is the same as the
parametrization proposed in the semiclassical theory of lightly
doped antiferromagnets by Shraiman and Siggia.\cite{ss}
As discussed in previous work \cite{rkk1,rkk2}, in the non-SDW
phase with $\langle z_\alpha \rangle = 0$, the parameterization in Eq.~(\ref{F})
and the coupling in Eq.~(\ref{eqSS}) lead to electron spectral
functions which are not centered at ${\bf K}_v$; once N\'eel order has
been disrupted, there is no special reason for the electronic
spectrum to be pinned at the magnetic Brillouin zone boundary.
The
computed \cite{rkk1} electron spectral functions have a
``Fermi arc'' structure, similar to those observed
experimentally. An additional mechanism for Fermi arc behavior is from
the phase fluctuations of the superconducting order,
and these effects will appear in our theory from the ``Josephson''
term introduced in Eq.~(\ref{Josephson}) between the $f_{\pm p}$ fermions
and the pairs formed out the states near ${\bf Q}_a$;  a recent work \cite{erez} has
examined classical thermal phase fluctuations present at high temperatuers,
and our formulation allows for a systematic consideration of quantum phase fluctuations
at low temperatures.

Indeed, our primary focus here is on the electronic excitations near the
${\bf Q}_a$ wavevectors. For these we need electromagnetic charge
$-e$ ``doublon'' annihilation operators $g_{q}$, where $q = \pm$.
The $g_\pm$ will be the central actors in our analysis. Note that
the $g_q$ do not carry any valley index, and the two charges of
$g_\pm$ specify all the fermionic degrees of freedom at all the
${\bf Q}_a$ in Fig.~\ref{fig:bz}. The $g_\pm$ are related to the
physical electrons by Eq.~(\ref{Psi}) and \cite{rkk3}
\begin{equation}
\left( \begin{array}{c} G_{1\uparrow} \\ G_{1\downarrow} \end{array} \right) = \mathcal{R}_z \left( \begin{array}{c} -g_- \\ -g_+ \end{array} \right)~~~~~,~~~~~\left( \begin{array}{c} G_{2\uparrow} \\ G_{2\downarrow} \end{array} \right) = \mathcal{R}_z \left( \begin{array}{c} g_- \\ -g_+ \end{array} \right),
\label{G}
\end{equation}
where the SU(2) rotation $\mathcal{R}_z$ was defined in Eq.~(\ref{Rz}).
In the SDW state, the $\pm$
indices of the $g_\pm$ fermions (and also of the $f_{\pm p}$ fermions) become equivalent to the $\uparrow,\downarrow$ spin indices quantized along the direction
of the N\'eel order {\em i.e.\/} the $g_\pm$ are conventional electron operators.~\cite{rkk1} However, in the phase with spin rotation
invariance preserved, $\pm$ gauge charges can be interpreted as sublattice indices which determine the sublattice on which
the fermion is predominantly (but not exclusively) located.

We will carry out our analysis in the framework of an effective field theory for the $g_{\pm}$ coupled
to the $A_\mu$ emergent gauge field.   The complete Lagrangian for our field theory has the following structure
(field theories for bosonic spinons and spinless fermions were also considered in early work \cite{wenholon,leeholon,shankar,ioffew}):
\begin{eqnarray}
\mathcal{L} &=& \mathcal{L}_g + \mathcal{L}_z +  \mathcal{L}_f  +  \mathcal{L}_{fg}  + \mathcal{L}_{zf} + \mathcal{L}_{zg} +\mathcal{L}_A \nonumber \\
\mathcal{L}_g &=& g_+^\dagger \left[ (\partial_\tau - i A_\tau + i e a_\tau)  - \frac{1}{2m^*} ( {\bm \nabla} - i {\bf A}  )^2 - \mu  \right] g_+
\nonumber \\ &+& g_-^\dagger \left[ (\partial_\tau + i A_\tau + i e a_\tau)  - \frac{1}{2m^*} ({\bm  \nabla} + i {\bf A} )^2 - \mu \right] g_-
\nonumber \\ &-&  \lambda g_+^\dagger g_-^\dagger g_- g_+ ~.\label{L}
\end{eqnarray}
We have only written out explicitly the Lagrangian $\mathcal{L}_g$ which involves the $g_{\pm}$ fermions, and which will be the
basis for almost all the computations in the body of this paper. The term $\mathcal{L}_{fg}$ coupling the $f_{\pm p}$ and $g_\pm$ fermions
will be described below; all other terms have been discussed previously \cite{rkk1,rkk3} and are recalled in Appendix~\ref{app:l}.
In Eq.~(\ref{L}), $a_\tau$ is the external electrostatic potential whose coupling shows that both $g_\pm$ carry charge $-e$.
The fluctuations of $a_\tau$ are controlled by the action
\begin{equation}
\mathcal{S}_a = \frac{1}{4 \pi} \int d \tau \int \frac{d^2 k}{4
\pi^2} \left|{\bf k}\right| \left|a_\tau ({\bf k}, \tau)\right|^2, \label{Sa}
\end{equation}
which leads to the Coulombic repulsion $e^2/r$ between all the $g_{\pm}$
particles. The magnetic dipole interactions associated with fluctuations of the electromagnetic vector potential
${\bf a}$ can be safely ignored.
The $g_\pm$ carriers have any effective mass $m^\ast$, and experience a chemical potential $\mu$.

\subsection{Fermion pairings}

\subsubsection{$g_\pm$ pairing}

We will be especially interested in the pairing of the $g_\pm$ fermions as described
by $\mathcal{L}_g$. Indeed, we will present arguments below in favor of the proposition that
an $s$-wave pairing of the $g_{\pm}$ is the primary pairing instability of the underdoped cuprates:
the pairing of the $f_{\pm p}$ fermions, and of the physical electrons $\Psi_{\alpha}$ will be shown to follow from it.

Eq.~(\ref{L}) already includes an attractive contact BCS interaction, $\lambda$,  between the $g_\pm$.
This attraction is permitted by
the underlying symmetries,~\cite{rkk3} and so can be written down on phenomenological grounds.
More physically, the longitudinal component of the $A_\mu$ gauge force provides an important component of the attractive
force between the $g_+$ and $g_-$ fermions: this is simply the attractive ``Coulomb'' force between
two opposite charges. This will be Thomas-Fermi screened by the compressible fermion state to an attractive force
with a range of order the Fermi wavelength.
It is clear that this force prefers an $s$-wave pairing between the $g_\pm$ fermions.
An additional
contribution to the $s$-wave attractive force comes from the term $\mathcal{L}_{zg}$ in Eq.~(\ref{lc2}).
Integrating out the $z_\alpha$ spinons, we find a contact attractive interaction $\sim -\lambda_{zg}^2$.

However, the key source of the $s$-wave pairing of the $g_\pm$
is the force associated with the {\em transverse\/} components of
the $A_\mu$ gauge field. As long as we are in the phase without
SDW order, these remain long-range and unscreened. The nature of
the $A_\mu$ fluctuations are similar to those of the fermionic
U(1) spin liquid~\cite{IL} or the Halperin-Lee-Read
state.~\cite{HLR} We have the following propagator for the
transverse part of the gauge field:
\begin{equation}
\label{AA} \left\langle A_i ({\bf q},\omega)  A_j (-{\bf q},
-\omega,)  \right\rangle = \left[\delta_{ij} - {q_i q_j \over q^2}
\right] \frac{1}{\chi q^2 + \gamma |\omega| /q  +\Delta_{\rm AF}
}.
\end{equation}
In our case, the effective gauge-field propagator contains
contributions both from the $z_\alpha$ spinons and from
the charge carrying fermions $f_{\pm p}$, $g_\pm$.
The spinon contributions to the susceptibility
$\chi$ from Eq.~(\ref{Lz}) were discussed in Ref.~\onlinecite{rkk4}.
The fermions yield $\chi = \left[ 6 \pi^2
\nu \right]^{-1}$, with the effective density of states, $\nu =
\overline{\mu}/\pi$, determined by the reduced mass of the holons
and doublons, $\overline{\mu} = {m^{\ast} m_f / (2m^{\ast} + m_f)}$
 (here $m_f$ is related to the masses in Eq.~(\ref{Lf}), and the factors of 2
arise from the valley degeneracy).
The damping term $\gamma$ comes from the
Landau damping of fermions and is given by the sum of two
Fermi-momenta $\gamma =\left(p_{\rm F}^{(g)} + 2p_{\rm F}^{\rm
(f)}\right)/(2\pi)$.
Finally, the ``mass'' term $\Delta_{\rm AF}$ arises from the Higgs
mechanism in the state with SDW order with
\begin{equation}
\Delta_{\rm AF}
\sim
\left|\langle z_\alpha \rangle \right|^2 ,
\end{equation}
as discussed in Appendix~\ref{app:l}.

Pairing due to transverse gauge forces has been considered
previously in the context of spin liquids. Because the magnetic
force between two oppositely directed currents is repulsive, a pairing between fermions 
could occur only in unusual channels,~\cite{Amper,KL} in particular in the ``Amperian'' channel where
the fermions on the same side of the Fermi surface pair up.~\cite{Amper} However, in
our case note that the $g_+$ and $g_-$ carry opposite $A_\mu$
gauge charges, and so the magnetic force is {\em attractive\/} in
the traditional $s$-wave BCS channel of pairing between fermions
on opposite sides of the Fermi surface. Indeed, this problem of
pairing by transverse gauge forces between Fermi surfaces of
opposite charges has been considered previously by Bonesteel,
McDonald, and Nayak,~\cite{bmn} and by Ussishkin and Stern
\cite{ady} in the context of double layer quantum Hall systems
each at filling fraction $\nu = 1/2$. In this quantum Hall
problem, the electrons in the two layers have opposite gauge
charges with respect to an ``antisymmetric'' U(1) gauge field
whose flux measures out-of-phase density fluctuations in the two
layers, and their $s$-wave BCS instability leads to a paired
quantum Hall state. An Eliashberg analysis of such a pairing
instability due to transverse gauge forces was carried out in
these works,~\cite{bmn,ady} and their results can be related to our
problem. An important result obtained in these studies was that
while the low-energy gauge fluctuations lead to very singular
electron self-energies in the normal state (including non-Fermi
liquid behavior), they are not \cite{rainer,msv} pair-breaking;
the pairing instability remains very strong. This should be
contrasted with the behavior near ferromagnetic quantum critical
points, where there is a similar anomalous self-energy in the
normal state, but the ferromagnetic fluctuations are pair-breaking
to $p$-wave superconductivity.~\cite{msv,roussev,cfhm} For our
problem, the estimate of the $s$-wave pairing temperature is
$T_{\rm p0} \sim {\gamma^2}/({m^{\ast3} \chi^2})$. Using the values
of $\chi$ and $\gamma$ quoted below Eq.~(\ref{AA}), and ignoring the spinon
contribution to $\chi$, we arrive at
the simple estimate $T_{\rm p0} \sim
E_{\rm F}$, where $E_{\rm F} = p_{\rm F}^2/(2 m^\ast)$
is the Fermi energy for electrons. To the extent
we can work within the context of $\mathcal{L}_g$ in
Eq.~(\ref{L}), we can understand this estimate
on dimensional grounds. Note that in the non-magnetic phase,
the only dimensional parameters appearing in Eq.~(\ref{AA}) are
associated with the Fermi surface, and there is no arbitrary
coupling constant in the coupling between $g_\pm$ and $A_\mu$. In
this respect, this problem is similar to the three-dimensional
Fermi gas at a Feshbach resonance. Consequently, the mean-field
pairing temperature $T_{\rm p0}$ can only be of order the available
energy scale, which are the Fermi energies. In reality, the actual
value of $T_{\rm p0}$ will be also influenced by the spinon contribution to $\chi$,
the Coulomb repulsion
$e^2/r$ between the $g_\pm$, the contribution of the $f_{\pm p}$ to
the $A_\mu$ polarization, and the value of $\lambda$.

Given the quenching of the transverse gauge propagator in
Eq.~(\ref{AA}) in the phase with $\langle z_\alpha \rangle \neq
0$, we can expect that the pairing instability will become weaker
in the SDW ordered state. This then sets up a natural and
appealing mechanism for the suppression of $T_{\rm p0}$ after the
onset of SDW order. Indeed, it offers a basis for the theory of
``competing orders''~\cite{demler} which has many attractive
phenomenological features.

\begin{figure}
\includegraphics[width=2.5in]{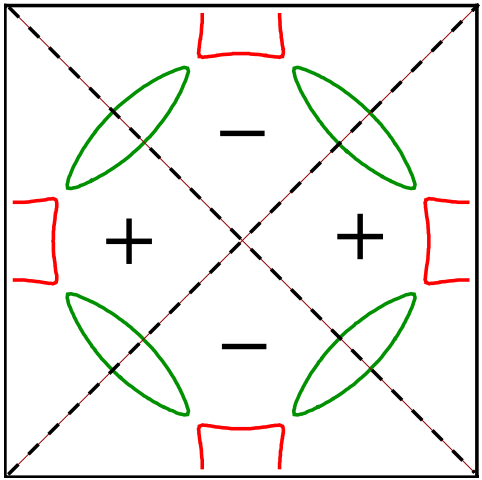}
\caption{Sign of the $d$-wave pairing amplitude superimposed on the electron and hole
pockets of the SDW metal. Note that these signs correspond to $s$-wave pairing of the electron pockets
and $p$-wave pairing of the hole pockets, as explained in the text.}
\label{fig:bz_sc}
\end{figure}

We have now established that $\mathcal{L}_g$ has a strong pairing instability to a state
where
\begin{equation}
\langle g_+ ({\bf k}) g_- (-{\bf k}) \rangle = \Delta_g,
\label{deltag}
\end{equation}
where we can take the pairing amplitude $\Delta_g$ to be independent of ${\bf k}$ near the Fermi level.
Then, what is the pairing amplitude for the physical electron operators in Eq.~(\ref{Psi}) ?
We assume, for simplicity, that we are in a non-magnetic state where $\langle z_{\alpha}^\ast z_\beta \rangle \sim
\delta_{\alpha\beta}$. Then from Eq.~(\ref{G}) we obtain
\begin{eqnarray}
&&\langle G_{1\alpha} ({\bf k}) G_{1 \beta} (- {\bf k}) \rangle = - \langle G_{2\alpha} ({\bf k}) G_{2 \beta} (- {\bf k}) \rangle
\sim \varepsilon_{\alpha\beta} \Delta_g; \nonumber \\
&&~~~~~~~~~~~~~~~~~~\langle G_{1\alpha} ({\bf k}) G_{2 \beta} (-
{\bf k}) \rangle = 0.
\end{eqnarray}
Comparing with Fig.~\ref{fig:bz}, we see that this is precisely the pairing signature expected for $d$-wave
pairing of the electrons; see also Fig.~\ref{fig:bz_sc}.

\subsubsection{$f_{\pm p}$ pairing}

Finally, we turn to the pairing of the $f_{\pm p}$ fermions. Just as
was the case for the $g_\pm$ fermions, the $A_\mu$ gauge forces
will prefer a $s$ wave pairing of oppositely charged $f_{\pm p}$
fermions.
However, there is a competing ``proximity'' effect
arising from the paired $g_\pm$ fermions. This proximity effect is
due to a Josephson coupling between $g_\pm$ and $f_{\pm p}$ pairs.
The form of such a Josephson term is tightly constrained by the
projective symmetry group (PSG), which was presented in detail in
Refs.~\onlinecite{rkk1,rkk3} and the transformations needed here
are listed in Table~\ref{psgtable}.
\begin{table}[t]
\begin{spacing}{1.5}
\centering
\begin{tabular}{||c||c|c|c||} \hline\hline
& $T_x$ &$R_{\pi/2}^{\rm dual}$ & $I_x^{\rm dual}$   \\
\hline\hline
$g_{+} g_{-}$ & $g_{+} g_{-}$ & $-g_{+} g_{-}$  & $g_{+} g_{-}$   \\
\hline
$f_{+1} f_{-1}$ & $-f_{+1} f_{-1}$ & $f_{+2} f_{-2}$ & $f_{+2} f_{-2}$ \\
\hline
$f_{+2} f_{-2}$ & $-f_{+2} f_{-2}$ & $-f_{+1} f_{-1}$ & $f_{+1} f_{-1}$ \\
 \hline
$f_{+1}  \stackrel{\scriptscriptstyle \leftrightarrow}{D}_x f_{-1}$ & $f_{+1}  \stackrel{\scriptscriptstyle \leftrightarrow}{D}_x f_{-1}$ & $-f_{+2}  \stackrel{\scriptscriptstyle \leftrightarrow}{D}_y f_{-2}$ & $f_{+2}  \stackrel{\scriptscriptstyle \leftrightarrow}{D}_x f_{-2}$ \\
\hline
 $f_{+1}  \stackrel{\scriptscriptstyle \leftrightarrow}{D}_y f_{-1}$ & $f_{+1}  \stackrel{\scriptscriptstyle \leftrightarrow}{D}_y f_{-1}$ & $f_{+2}  \stackrel{\scriptscriptstyle \leftrightarrow}{D}_x f_{-2}$ & $-f_{+2}  \stackrel{\scriptscriptstyle \leftrightarrow}{D}_y f_{-2}$ \\
  \hline
$f_{+2} \stackrel{\scriptscriptstyle \leftrightarrow}{D}_x f_{-2}$ & $f_{+2}  \stackrel{\scriptscriptstyle \leftrightarrow}{D}_x f_{-2}$ & $f_{+1}  \stackrel{\scriptscriptstyle \leftrightarrow}{D}_y f_{-1}$ & $f_{+1}  \stackrel{\scriptscriptstyle \leftrightarrow}{D}_x f_{-1}$ \\
  \hline
$f_{+2} \stackrel{\scriptscriptstyle \leftrightarrow}{D}_y f_{-2}$ & $f_{+2}  \stackrel{\scriptscriptstyle \leftrightarrow}{D}_y f_{-2}$ & $- f_{+1}  \stackrel{\scriptscriptstyle \leftrightarrow}{D}_x f_{-1}$ & $-f_{+1}  \stackrel{\scriptscriptstyle \leftrightarrow}{D}_y f_{-1}$ \\
\hline\hline
\end{tabular}
\end{spacing}
\caption{PSG transformations deduced from the PSG of the $f_{\pm q}$ in Table II in Ref.~\onlinecite{rkk1},
the PSG of the $g_\pm$ in Table III in Ref.~\onlinecite{rkk3}, and the PSG of ${\bf A}$ in Eq.~(12) of Ref.~\onlinecite{annals}.
The transformations are
$T_x$: translation by one lattice
spacing along the $x$ direction; $R_{\pi/2}^{\rm dual}$: 90$^\circ$
rotation about a dual lattice site on the plaquette center
($x\rightarrow y,y\rightarrow-x$); $I_x^{\rm dual}$: reflection
about the dual lattice $y$ axis ($x\rightarrow -x,y\rightarrow y$).}
\label{psgtable}
\end{table}
An analysis based on Table~\ref{psgtable} shows that
the simplest allowed coupling between the $g_\pm$ and $f_{\pm p}$ pairs which is invariant under the PSG is
\begin{eqnarray}
\mathcal{L}_{fg} = -i J_{fg} \, \Bigl[ g_+  g_- \Bigr] \, \Bigl[
f_{+1} \stackrel{\scriptscriptstyle \leftrightarrow}{D}_x f_{-1} - f_{+1} \stackrel{\scriptscriptstyle \leftrightarrow}{D}_y f_{-1}
+ f_{+2} \stackrel{\scriptscriptstyle \leftrightarrow}{D}_x f_{-2} +  f_{+2} \stackrel{\scriptscriptstyle \leftrightarrow}{D}_y f_{-2}  \Bigr] + \mbox{H.c.},
\label{Josephson}
\end{eqnarray}
where $J_{fg}$ is the Josephson coupling, $D_i \equiv \partial_i - q A_i$ is the co-variant derivative acting on a field
with charge $q$,  and
$a \stackrel{\scriptscriptstyle \leftrightarrow}{D}_i b  \equiv a D_i b - (D_i a) b$.
Note that Table~\ref{psgtable} does not permit any term without a spatial gradient.
From the structure of the Josephson coupling in Eq.~(\ref{Josephson}) we see that the proximity
effect induces a $p$-wave pairing of the $f_{\pm p}$ fermions. Thus there is {\em frustration\/} in the $f_{\pm p}$
pairing, with the $A_\mu$ gauge forces preferring $s$-wave.

At the microscopic level, a computation of the pairing of the $f_{\pm p}$ fermions in the SDW ordered state
has been carried out by Sushkov and collaborators.~\cite{sushkov0,sushkov1,sushkov2,sushkov3} They showed that,
for a suitable range of parameters, the long-range spin-wave interaction preferred a $p$-wave pairing. In our formulation
this long-range attraction between the $f_{\pm p}$ is mediated by the Shraiman-Siggia term in Eq.~(\ref{eqSS}).
Note that this Shraiman-Siggia term does not apply to the $g_\pm$, and so the corresponding interaction is absent there.

On the basis of our arguments above, and the complementary microscopic computations \cite{sushkov0,sushkov1,sushkov2,sushkov3},
we conclude that the $s$-wave pairing of the $g_\pm$
fermions is the dominant instability, and the Josephson coupling in Eq.~(\ref{Josephson})
induces a sympathetic $p$-wave pairing of the $f_{\pm p}$ fermions. A key point is that the $A_\mu$ gauge forces
will be pair-breaking towards this $p$-wave pairing,~\cite{msv,roussev,cfhm} and consequently the $f_{\pm p}$
pairing amplitude will be quite weak.

Specifically,
combining Eqs.~(\ref{Josephson}) and (\ref{deltag}), we deduce that the proximity-effect pairing of the $f_{\pm p}$
fermions induced by the $g_{\pm}$ fermions has the $p$-wave form
\begin{eqnarray}
\langle f_{+1} ({\bf k}) f_{-1} (-{\bf k}) \rangle  & \sim & (k_x - k_y) J_{fg}  \Delta_g; \nonumber \\
\langle f_{+2} ({\bf k}) f_{-2} (-{\bf k}) \rangle  & \sim & (k_x + k_y) J_{fg} \Delta_g;  \nonumber \\
\langle f_{+1} ({\bf k}) f_{-2} (-{\bf k}) \rangle  & = & 0,
\end{eqnarray}
where the momentum dependencies are a consequence of the spatial gradients in Eq.~(\ref{Josephson}),
and the pairing amplitudes between fermions with like $A_\mu$ charges are zero.
Finally, from these results and Eq.~(\ref{F}), we can deduce the pairing
of the physical electron operators in $F_{v\alpha}$ in Eq.~(\ref{Psi}):
\begin{eqnarray}
\langle F_{1\alpha} ({\bf k}) F_{3 \beta} (-{\bf k}) \rangle  & \sim & \varepsilon_{\alpha\beta} (k_x - k_y) J_{fg}  \Delta_g; \nonumber \\
\langle F_{2\alpha} ({\bf k}) F_{4 \beta} (-{\bf k}) \rangle   & \sim & \varepsilon_{\alpha\beta} (k_x + k_y) J_{fg} \Delta_g
\label{Fpair}
\end{eqnarray}
and all other $F_{v\alpha}$ pairings vanish. A glance at
Figs.~\ref{fig:bz} and ~\ref{fig:bz_sc} shows that these are precisely the pairings
associated with a $d$-wave pairing signature of the physical
$\Psi_\alpha$ electrons.  The momentum dependencies in
Eq.~(\ref{Fpair}) shows that the pairing amplitude changes sign
across the Brillouin zone diagonals. Also, the vanishing of the
pairing along these zone diagonals shows that there will be
gapless ``nodal'' fermionic excitations.

We close this section by noting that the structure of the Josephson coupling in Eq.~(\ref{Josephson}) is closely connected
to that appearing in the model of Geshkenbein, Ioffe and Larkin.~\cite{geshkenbein}
They
considered a phenomenological model of charge $-2e$ bosons, representing pairs of electrons near the ${\bf Q}_a$, coupled to fermions near
the ${\bf K}_v$. Their boson-fermion Josephson coupling had a matrix element which changed sign along the Brillouin zone
diagonals.
Identifying their boson $b$ as $b \sim g_+ g_-$, we see that Eq.~(\ref{Josephson}) also shares these features.

\section{Quantum oscillations in the normal state}
\label{sec:qosc}

This section will consider the low temperature transport properties of the paired
state of the $g_\pm$ described by $\mathcal{L}_g$ which is driven normal by a strong
applied magnetic field, $H$.

We begin by a simple estimate of the depairing field, $H_{\rm p2} (0)$, associated to the pairing temperature,
$T_{\rm p0}$. In the absence of a complete Eliashberg theory of the influence
of the transverse gauge fluctuations, we will be satisfied here with an estimate based on the weak-coupling BCS theory result for 
the upper critical field of a clean two-dimensional superconductor (note that the value of the upper critical field depends on the purity~\cite{HW,GHc2} and dimensionality of the system~\cite{MinJap,GaL}). For the purpose of numerical estimates, we ignore here the quantum oscillation phenomena in the transition point itself, which as discussed below may lead to a reentrant behavior. Within these assumptions, we have the following BCS formula~\cite{HW,GaL} that we associate with the ``quantum depairing field''  (here, we restore the fundamental physical constants):
\begin{equation}
\label{Hc2}
{e H_{\rm p2}(0) \over m^{\ast} c} = {\pi^2 \over \gamma_E} {k_{\rm B} \over \hbar}   {T_{\rm p0}^2 \over T_{\rm F}},
\end{equation}
where $e$ is the electron charge, $c$ is the speed of light, $m_*$ is the effective mass of carriers in the pocket,
$k_{\rm B}$ is the Boltzmann constant, $T_{\rm F} = p_{\rm F}^2/(2 k_{\rm B} m^*)$ is the Fermi temperature
for the electrons in the pocket, $p_{\rm F}$ is their Fermi momentum, and $\gamma_E = 1.781\dots$ is the
exponential of Euler's constant. The onset of quantum oscillations
in the cuprate experiments\cite{doiron,cooper,nigel,cyril,suchitra} is identified here with
the quantum depairing field, $H_{\rm p2}(0)$. The former is about
$H_{\rm p2}(0) \sim 50~{\rm Tesla}$. The quantum oscillation
measurements also provide information for the effective electron
mass, which appears to be of order free electron mass or a few times larger
(the exact values vary in experiment) and for the area of the electron Fermi
surface,  which is estimated to be a few percent of the total area of
the Brillouin zone, which in turn is determined by the lattice
constants for YBCO. This information allows us to extract all
necessary parameters. The Fermi temperature is related to the frequency
of quantum oscillations, $F_{\rm SdH}$, as follows:
\begin{equation}
\label{TF}
T_{\rm F} = {\pi  \hbar^2 \over k_{\rm B} m_e \Phi_0} \left({m_e \over m^{\ast}}\right)
F_{\rm SdH},
\end{equation}
where $\Phi_0 = (2.07 \times 10^{-15})\,{\rm T}\cdot{\rm m}^2$ is the flux quantum. The first factor
in Eq.~(\ref{TF}) contains
only fundamental  constants and is equal to
${\pi  \hbar^2 \over k_{\rm B} m_e \Phi_0} = 1.33\,{\rm K}/{\rm T}$ (${\rm K}$ and ${\rm T}$
correspond  to  the units of Kelvin and Tesla respectively). Using Eq.~(\ref{Hc2}), we can write the following
relation between the pairing temperature and the quantum pairing field and the Fermi temperature:
\begin{equation}
\label{Tp0Hc2}
T_{\rm p0} = \sqrt{ {\gamma_{\rm E} \over \pi^2} {\mu_{\rm B} \over k_{\rm B}} \left({2m_e \over m^{\ast}}\right)
T_{\rm F} H_{\rm p2}(0)},
\end{equation}
where $\mu_{\rm B} = e\hbar/(2|e|c)$ is the Bohr's magneton.
Converting all quantities to the units of Tesla and Kelvin relevant to the interpretation of experimental data and
using the actual values of the corresponding physical constants, we can express the Fermi temperature
for the electrons in the pocket and the corresponding pairing temperature as follows
\begin{equation}
\label{TFe}
T_{\rm F}~\mbox{K} \approx \left( {1.33 m_e \over m^{\ast}} \right)
[F_{\rm SdH}~\mbox{T}]
\end{equation}
and
\begin{equation}
\label{Est}
T_{\rm p0}~\mbox{K} \approx \sqrt{ {0.24 m_e \over m^{\ast}}\,  [T_{\rm F}~\mbox{K}][H_{\rm p2}(0)~\mbox{T}] },
\end{equation}
where $[T_{\rm F}~\mbox{K}]$ is the Fermi temperature expressed in Kelvin,
and $[H_{\rm p2}(0)~\mbox{T}]$ and $[F_{\rm SdH}~\mbox{T}]$ are the quantum critical
field  and the period of the Shubnikov-de Haas
oscillations expressed in Tesla. We emphasize that the above estimates assume the applicability
of the weak-coupling BCS theory and a circular Fermi surface for the electron pocket.
Therefore, Eqs.~(\ref{Hc2} -- \ref{Est}) are not expected to determine exactly the
numerical coefficients, but should provide  correct order-of-magnitude estimates. Using the
available experimental data, {\em e.g.} from Ref.~[\onlinecite{doiron}], which estimates
the period of oscillations to be $F_{\rm SdH} \sim 530~\mbox{T}$ and
onset of oscillations (the critical pairing field in our theory) as $H_{\rm p2}(0) \sim 50~\mbox{T}$,
we get the following relation between the zero-field pairing temperature and the Fermi
temperature:~$T_{\rm p0}~[\mbox{K}] \approx
\sqrt{m_e / m^{\ast}}\,  \sqrt{12~[\mbox{K}]\,[T_{\rm F}~\mbox{K}]}$ and
$T_{F} \approx \left({m_e / m^{\ast}}\right)~700~\mbox{K}$. Finally, the estimates for
the actual numerical values of $T_{\rm F}$ and $T_{\rm p0}$ depend on the effective mass for electrons.
Various experiments report different values for the latter, $m^{\ast} \sim (1 \mbox{ --- } 3) m_e$.
As explained below in Secs.~\ref{sec:mirlin}
and~\ref{sec:al}, one should be careful in extracting the effective mass from the temperature dependence of
the oscillations in this phase, because there may be other effects due to superconducting and gauge fluctuations,
which will change the temperature dependence of the amplitude in the Lifshitz-Kosevich formula.
In addition, if the Fermi surface and/or the quasiparticle weight factor are anisotropic, it would also
modify the effective temperature dependence in the Lifshitz-Kosevich formula~\cite{Zosc}
In particular, if the anisotropy of the quasiparticle renormalization $Z$-factor, $Z_{\bf p}$,
is not taken into account, the effective mass extracted from the quantum oscillation measurements
will overestimate the actual effective mass by the factor of $\left\langle 1/ Z_{\bf p} \right\rangle_{\rm FS}$,~\cite{Zosc}
where the angular brackets imply averaging over the Fermi surface. However, if we now assume that the
effective mass for the electronic excitations in the pocket is of order
free electron mass (as suggested by experiments), $ m^{\ast} \sim m_e$, then we get the Fermi temperature
of the electron pocket $T_{F} \approx 700~\mbox{K}$, the pairing temperature $T_{\rm p0} \approx 100~\mbox{K}$,
and the zero-temperature BCS superconducting gap,
$\Delta/k_{\rm B} = \left({\pi   / \gamma_{\rm E}}\right) T_{\rm p0} \approx 200~\mbox{K}$.

The relatively large ratio between the electron pairing temperature and the Fermi temperature
$\left(T_{\rm p0}/T_{\rm F}\right) \sim (1/7)$ justifies our earlier conclusion about
strong Cooper pairing in the electron pocket.
We note further that a complete description of the finite temperature pseudogap region likely requires the inclusion
of further interactions between the Cooper pairs. A particularly interesting possibility appears within an effective model where the
paired electrons in the pockets interact on the lattice. This type of model for the electron pairs, $\left\langle g_+ g_- \right\rangle$,
may have a superconductor-to-insulator phase transition with a Mott-type gap of order, $J$, accompanied by development
of charge order associated with the density of bosons \cite{bbbss} and the monopole
Berry phases \cite{bbbss2,annals} in $\mathcal{L}_A$
(see Appendix~\ref{app:gauge}).

In the following, we will describe the Shubnikov-de Haas
oscillations in the resistivity at $H>H_{\rm p2} (T)$. For reference,
we recall the Lifshitz-Kosevich formula for the oscillatory
component of the resistivity $\rho$  (retaining only the lowest
oscillation harmonic)
\begin{equation}
\label{LK} {\rho_{\rm osc}(H) \over \rho_{||} (H=0)} = { X(T) \over
\sinh \left[ X(T) \right]} \exp \left(-\frac{\pi}{\omega_c \tau}
\right) \cos\left( {2\pi E_{\rm F} \over \omega_c} \right),
\end{equation}
where $\rho_{||}(H=0)$ is the Drude resistivity in zero field,
$X(T) = 2 \pi^2 T/\omega_c$, $\tau$ is the elastic scattering time
from impurities, $\omega_c = e H/(m^\ast c)$ is the cyclotron
frequency.

We begin our analysis of transport by discussing the fate of the Ioffe-Larkin
composition rule in our system in Section~\ref{sec:ilc}.
In Section~\ref{sec:mirlin}, we will describe the corrections to
the Lifshitz-Kosevich result from the fluctuations of the $A_\mu$
gauge field, while the influence of pairing fluctuations will be
discussed in Section~\ref{sec:al}.

\subsection{Ioffe-Larkin composition rule}
\label{sec:ilc}

Before turning to the computation of the transport properties at $H>H_{\rm p2} (T)$ in the subsections below,
we need to discuss an important, but technical,
issue. In previous studies of spin-charge separation in the cuprates, a crucial ingredient in the computation
of the physical conductivity was the Ioffe-Larkin composition rule.~\cite{IL,KI} This states the resistivities of the
spinons and charge carriers add to yield the physical resistivity. In our present situation, there is a crucial difference
from the models considered in these works: our $g_\pm$ fermions carry opposite charges under the internal
$A_\mu$ gauge field, and the same charge under the electromagnetic gauge field $a_\mu$. In contrast, the previous
theories had holons carrying only a single charge under the analog of $A_\mu$. An immediate consequence for our theory
is that the cross-polarization operator between the two gauge fields, $\Pi_{Aa}$, vanishes identically:
the $g_+$ and $g_-$ fermions induce opposite polarizations which cancel each other.
(More formally, this can be seen by the PSG of the ${\bf A}$ gauge field \cite{annals}, which
changes sign under translation by a lattice site, and so cannot couple linearly to ${\bf a}$ at long wavelengths.)
In other words, in the presence
of an applied electromagnetic field $a_\mu$, the physical
current is carried equally by the $g_+$ and the $g_-$. In this current carrying
state, the $A_\mu$ currents of the $g_+$ and $g_-$ travel in opposite directions, leading to no net
internal gauge current; consequently, there is also no spinon current.
The final conclusion is then very simple: the physical conductivity is just the sum of the conductivities of the $g_+$
and $g_-$, and the traditional Ioffe-Larkin rule does not apply to our model.

\subsection{Gauge field fluctuations}
\label{sec:mirlin}

The influence of gauge fluctuations on magnetotransport was examined in the context of
the $\nu=1/2$ quantum Hall state,~\cite{mirlin,mirlin2} and here we will adapt these earlier
results to our problem. This analysis was carried out using the quasiclassical method, in which
the gauge field fluctuations are treated as a random static ``magnetic'' field which influences the cyclotron
motion of the fermions. We will follow the same method here.

There are two potential sources of the random field. In the quantum Hall case, the most important
source was the local field induced by the Chern-Simons term from the impurity potential.
This source is absent in our case, as we do not have a Chern-Simons term.
Indeed, the PSG of the ${\bf A}$ field \cite{annals} shows that only impurities which locally break
time-reversal can induce a non-zero flux of ${\bf A}$; we will assume that such impurities are absent.
An important consequence is that the amplitude of the SdH oscillations in Eq.~(\ref{LK})
remains unaffected at $T=0$ by the presence of the gauge field.

The second source of the random field was the thermal fluctuations of ${\bf A}$. For a random
field, $h = {\bm \nabla} \times {\bf A}$, with equal time correlations given by
\begin{equation}
\langle h ( {\bf r}) h ({\bf r}') \rangle = U (|{\bf r} - {\bf r}'|),
\end{equation}
Mirlin {\em et al.}~\cite{mirlin2} showed that the SdH oscillations in Eq.~(\ref{LK})
are suppressed by a factor $\exp ( - S_h )$ where
\begin{equation}
S_h = \pi R_c^2 \int_0^\infty \frac{dq}{q} J_1^2 (q R_c) \widetilde{U}(q).
\label{sh1}
\end{equation}
Here $R_c = \sqrt{2 E_{\rm F}/m^\ast} /\omega_c$ is the cyclotron radius of fermions at the Fermi
level, and $\widetilde{U} ({\bf q})$ is the Fourier
transform of $U({\bf r})$. In the quasiclassical limit, the equal-time gauge field correlations
can be evaluated from Eq.~(\ref{AA}) to yield
\begin{equation}
\widetilde{U} (q) = \frac{T q^2}{\chi q^2 + \Delta_{\rm AF}}.
\label{sh2}
\end{equation}
We can also deduce from Eq.~(\ref{AA}) a necessary condition for
the applicability of the quasiclassical approximation: the
characteristic frequency $\omega \sim (q/\gamma) (\chi q^2 +
\Delta_{\rm AF})$ at the characteristic wave-vector $q \sim 1/R_c$
should be smaller than $T$. Now, we can  insert Eq.~(\ref{sh2})
into Eq.~(\ref{sh1}) and obtain
\begin{equation}
S_h = \frac{ \pi R_c^2 T}{ \chi} I_1 \left( R_c
\sqrt{ {\Delta_{\rm AF}}/{\chi}} \right)  K_1 \left( R_c
\sqrt{ {\Delta_{\rm AF}}/{\chi}} \right),
\label{sh3}
\end{equation}
where $I_1$ and $K_1$ are modified Bessel functions.
In the phase without SDW order, where $\Delta_{\rm AF} = 0$, we then have
\begin{equation}
S_h = \frac{\pi E_{\rm F} T}{ m^\ast \chi \omega_c^2}.
\label{sh4}
\end{equation}
The value of $S_h$ decreases monotonically
into the phase with SDW order {\em i.e.\/} the SdH oscillations have a larger amplitude in the SDW state.
Deep in the SDW state, where $\Delta_{\rm AF} \gg \chi/R_c^2$, we have the limiting result for $S_h$
(which is always smaller than the value of $S_h$ in Eq.~(\ref{sh4}))
\begin{equation}
S_h = \frac{\pi T}{\omega_c} \sqrt{ \frac{ E_{\rm F} }{ 2 m^\ast
\chi \Delta_{\rm AF}}}. \label{sh5}
\end{equation}
The thermal suppression of the SdH oscillations in Eqs.~(\ref{sh3}-\ref{sh5}) by the factor
$\exp (- S_h ) $ will combine with the factor $\exp (- 2 \pi^2 T/\omega_c)$ already present in Eq.~(\ref{LK}).
While the $T$ dependencies in the two factors are the same, they are distinguished by their $B$ dependencies.
In particular, we have $S_h \sim T/B^2$ in Eq.~(\ref{sh4}), and this can serve as a characteristic signature
of gauge field fluctuations in an algebraic charge liquid.

\subsection{Pairing fluctuations}
\label{sec:al}

This subsection will describe the corrections to Eq.~(\ref{LK}) in the context of a traditional
fluctuating superconductivity computation built on BCS theory. We will not examine the interesting
question of how the transverse $A_\mu$ fluctuations will modify the Cooperon operator. However, given
the absence of pair-breaking effects in the Eliashberg computation,~\cite{bmn,ady} it is reasonable to
expect that the Cooperon will remain the same near the pair-breaking transition. In any case, we can
also appeal to the onset of SDW order, which leads to $\Delta_{\rm AF} > 0$, to quench the gauge
fluctuations.

We begin with the Cooperon
operator in the quasiclassical approximation as follows
\begin{equation}
\label{C12}
C ({\varepsilon}, \omega; {\bf r},{\bf r}') =  {\cal G}_{\varepsilon +\omega} ({\bf r} - {\bf r}')
{\cal G}_{-\varepsilon} ({\bf r} - {\bf r}') \exp \left[ -2ie \int_{\bf r}^{\bf r'} {\bf a} \cdot d{\bf l}\right],
\end{equation}
where ${\cal G}_{\varepsilon} ({\bf  \rho})$ is the fermion Green's function in the absence of a magnetic field,
and the latter (real magnetic field), ${\bf H} = {\bm \nabla} \times {\bf a}$, is accounted for only in the gauge
factor. Note that in our model, the $\pm$-electrons carry opposite $e_* = \pm 1$ charges with respect
to the ``internal'' gauge field, ${\bf A}$, but have the same  (negative) electron charge, $-e$, with respect to the
external electromagnetic field. Hence, we can take advantage of the old results of Helfand and Werthamer,~\cite{HW}
 who have
proven that the Cooperon operator, $\hat{C}(\varepsilon,\omega)$, whose kernel is defined via Eq.~(\ref{C12}), is a diagonal operator
in the Landau basis. Its matrix elements are defined simply by the expression without magnetic field,
$C_n = \left\langle n~\left| (\varepsilon, \omega;{\bf q} \to \hat{\bm \pi} ) \right|~n\right\rangle$, but
with the momentum ${\bf q}$ replaced with the operator of the kinetic momentum of a Cooper pair
$\hat{\bm \pi} = \left[ {\bf q} - 2ie{\bf a} (\hat{\bf r})\right]$. The corresponding matrix elements are known
from the Landau problem in the elementary single-particle quantum mechanics, {\em e.g.},
$\left\langle n\left| \hat{\bm \pi}^2  \right|n\right\rangle = 4 e H (n+1/2)$
(note that the Cooper pair charge is $-2e$
and mass is $2m^{\ast}$). The general expression for the Cooperon  without a magnetic field is as follows:
\begin{equation}
\label{bubble}
C( \varepsilon, \omega; {\bf q})=
2 \pi \nu
{ \theta \left[\varepsilon \left( \varepsilon - \omega \right) \right] \over
\sqrt{
\left( 2 \varepsilon - \omega + {(1 / \tau)} {\rm sgn \,}\varepsilon \right)^2 + v_F^2 q^2}},
\end{equation}
where $\nu$ is the density of states at the Fermi level and $\tau$ is the scattering time. Since
we are interested in explaining the quantum oscillations, we assume that the latter is large and
set it to $\tau  =\infty$. Note that the clean case is in fact more complicated
than the disordered limit, because the Green's functions and the Cooperon are non-local
objects ({\em i.e.\/}, there is no exponential decay in space).
The fluctuation propagator
 for superconducting fluctuations is an operator given by
\begin{equation}
\label{OpL}
\hat{\cal L} (\omega) = \left[ \lambda_{\rm eff}^{-1} - T \sum\limits_\varepsilon \hat{C}( \varepsilon, \omega)
\right]^{-1}.
\end{equation}
For the purpose of describing quantum oscillations, we are
interested only in the quantum critical point, $H_{\rm p2}(0)$, which
is determined by the divergence of the matrix element at the
lowest Landau level of  the operator, $\overline{\cal L}_0 (0) = \infty$.
This leads to the expression near the quantum pairing field as follows
(here and below, the index ``0'' corresponds to the matrix element at the lowest Landau level):
\begin{equation}
\label{L1} \overline{\cal L}_0 (\omega) = -{1\over \nu} \left[  r
+ \sqrt{\gamma \over \pi} {|\omega| \over T_{\rm p0}} \right]^{-1},
\end{equation}
where $r = \left[ H-H_{\rm p2}(0) \right]/H$ is the proximity
to the pairing transition, and the value of the critical field
was specified in Eq.~(\ref{Hc2}).

In the expressions so far, the Cooperon dependence on the magnetic
field is accounted for only via the gauge factor (\ref{C12}).
Physically this corresponds to an approximation in which the
motion of the Cooper pair in the magnetic field is quantized (more
precisely, the center-of-mass motion is quantized), but the
cyclotron motion of individual electrons within a Cooper pair is
not accounted for. This quasiclassical approximation is valid if
either temperature is relatively large, $T \gg \omega_c$ (note
that near $H_{\rm p2}(0)$, $\omega_{\rm p2}/T_{\rm p0} \sim T_{\rm p0}/E_{\rm F}$,
which is small in the conventional weak-coupling BCS theory), or
if disorder is strong enough, $\omega_c \tau \ll 1$. In the
regime, where the oscillations are observed neither of these
conditions is satisfied and therefore one has to take into account
Landau quantization of electrons within a Cooper pair. This
problem was considered back in the sixties, {\em e.g.\/}, by
Gruenberg and G\"unter,~\cite{GG} and we reiterate here the main
steps to derive the oscillating transition point and the
fluctuation effects in its vicinity. The quantity of interest is
the fluctuation propagator, which we write as:
\begin{equation}
\label{L2} {\cal L}_0 (\omega) = {1 \over \overline{\cal
L}_0(\omega) - C_{\rm osc} },
\end{equation}
where $ \overline{\cal L}_0(\omega)$ is the fluctuation propagator given by Eq.~(\ref{L1}), which does not
take into account oscillations, and $C_{\rm osc}$ is the correction to the Cooperon with the quantum
oscillation effects, given by
\begin{equation}
\label{Cosc1} C_{\rm osc}  = \int e^{-eHr^2/4}\,C({\bf r}) d^2 r -
\overline{C}_0,
\end{equation}
with $\overline{C}_0$ is the matrix element for the Cooperon at the lowest Landau level without oscillations
and $C({\bf r}) = T \sum\limits_{\varepsilon} {\cal G} (\varepsilon,{\bf r};B)
{\cal G} (-\varepsilon,{\bf r}; B)$ and the electron Green's function in a magnetic field is given by
\begin{equation}
\label{G(B)} {\cal G} (\varepsilon,{\bf r}; B)  = \nu_{\rm mag}
\omega_c \sum\limits_{n_e=1}^\infty {e^{-{eHr^2 / 4}}
L_{n_e}\left({eHr^2 / 2}\right) \over i\varepsilon -
\omega_c\left( n_e + {1/ 2} \right) - \mu + i{\rm sgn}\,
(\varepsilon)/(2\tau)},
\end{equation}
with $L_{n_e}(z)$ being the Laguerre polynomial at the $n_e$-th
{\em single-electron} Landau level. It is these electron Landau
levels that may generate de Haas oscillations above and even
within the superconducting phase. The quantity $C_{\rm osc}$ has
been considered previously by Mineev~\cite{Min99} and also by
Larkin and one of the authors,~\cite{GaL} and it has the following
form (again, we retain only the leading oscillation term, and drop
all the higher-order harmonics)
\begin{equation}
\label{Cosc}
C_{\rm osc}(H)  = {4 \nu \over 3 \sqrt{\pi}} \sqrt{\omega_c \over E_{\rm F}} \,
{3 X(T)
\over \sinh \left[ 3 X(T) \right]}\, \exp\left(-{3\pi \over \omega_c \tau} \right) \cos\left( {2\pi E_{\rm F} \over \omega_c} \right),
\end{equation}
where $X(T) = 2\pi^2T/\omega_c$ and $\pi/ (\omega_c \tau)$  are the familiar terms,
which describe the suppression of quantum oscillations by the temperature and disorder (Dingle factor)
correspondingly. Note however that there is an additional factor of 3 in these suppression terms in
 the leading oscillation harmonics for the Cooperon. This additional suppression
 (first pointed out by Mineev~\cite{Min99,Min})
 is due to the fact that to resolve quantum oscillations coming out of a Cooper pair built of two electrons,
 one has to ``resolve'' their relative cyclotron motion without breaking the pair.

 Using the expression (\ref{Cosc}),  we obtain the fluctuation
 propagators follows
\begin{equation}
\label{L3} {\cal L}_0 (\omega) = -{1 \over \nu} \left[  {H -
\overline{H}_{\rm p2}(T) \over H} - C_{\rm osc}/\nu + \sqrt{\gamma
\over \pi} {|\omega| \over T_{\rm p0}} \right]^{-1}.
\end{equation}
{}From Eq.~(\ref{L3}), we see that the oscillatory part in the
Cooperon can be interpreted as a correction to the upper critical
field, which too may oscillate and therefore pairing may show a
re-entrant behavior at low temperatures. Hence in the regime where
quantum oscillations are observed, the exact value of the ``upper
critical field'' (even in the sense of the BCS pairing-depairing
transition) is strictly-speaking ill-defined because there are
many critical fields as long as oscillations are not suppressed.

Another important circumstance has been pointed out by Champel and
Mineev,~\cite{Min} who argued that even below the mean-field
critical field, $\overline{H}_{\rm p2}(T)$, where the system is
paired, one may see (de Haas-van Alfven) oscillations in the
gapless superconductivity region, which is determined by the
condition $\left[H_{\rm p2}(0)  - H\right]/ H \ll \sqrt{\omega_{\rm p2} / E_{\rm F}} \ln
({\omega_{\rm p2} / E_{\rm F}})$ in three dimensions and $\left[H_{\rm p2}(0)  -
H\right]/ H \ll \sqrt{\omega_{\rm p2} / E_{\rm F}}$ in strictly two dimensions [here,
$\omega_{\rm p2} = eH_{\rm p2}(0)/(m^{\ast} c)$].
We note that $\sqrt{\omega_{\rm p2} / E_{\rm F}}$ is the Ginzburg
parameter, which is typically negligibly small in the conventional
BCS systems, but is expected to be larger in the cuprates.
The numerical estimates (\ref{Hc2} -- \ref{Est}) in the beginning
of this section, suggest a very wide fluctuation Ginzburg region for the
strongly-paired small electron pocket of our model; {\em E.g.\/}, using
the experimental data of Ref.~[\onlinecite{doiron}], we get
$\sqrt{\omega_{\rm p2} / E_{\rm F}} \sim 1/3$ for the electron pocket \cite{ginzburg}.
We also emphasize that these possible quantum
oscillations in the gapless superconductivity region are different
from the effect, which may arise from the normal vortex cores well
below the critical field. In fact, the latter effect may be
significantly suppressed in the strongly paired phase, where the
individual vortex cores are not large enough to support a
many-body electron state leading to quantum oscillations.

\begin{figure}[h]
 \includegraphics[width=5in]{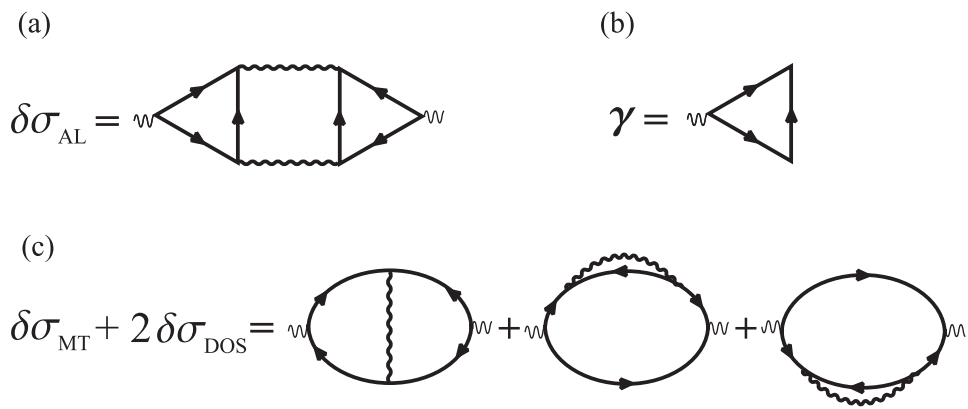}
 \caption{The relevant diagrams that describe the fluctuation transport near a pairing transition. Fig~(1a) shows the Aslamazov-Larkin
 contribution to conductivity in the clean limit; Fig~(1b) shows the current vertex of the Aslamazov-Larkin diagram
 calculated in Eqs.~(\protect\ref{cvert}) and (\protect\ref{vot}); Fig~(1c) describes the combined contribution
 of the Maki-Thompson  and density-of-states terms. }\label{fig:diag}
\end{figure}

We now ready to calculate the fluctuation corrections to the
Lifshitz-Kosevich formula in Eq.~(\ref{LK}). The fluctuation
conductivity in the Gaussian approximation is given by the sum of
the Aslamazov-Larkin, density-of-states, and Maki-Thompson
diagrams~\cite{LV} (see Fig.~\ref{fig:diag}). In a strong magnetic
field and low temperature, all these diagrams are generally of the
same order and in a disordered system all play an important role.
However, in the super-clean case of $\omega_c \tau \to \infty$,
the Maki-Thompson  and density-of-states terms cancel each other
exactly at least for the non-oscillating terms~\cite{Varlam,GaL}
and only the Aslamazov-Larkin term~\cite{AL} survives. The
longitudinal part of the electromagnetic response tensor has been
considered by Larkin and one of the authors~\cite{GaL} and reads
\begin{equation}
\label{Qcl}
Q(\omega) =8 \nu e^2 T \sum\limits_{\Omega}
\gamma^2_{01}(\Omega,\omega) {\cal L}_0(\Omega) {\cal L}_1,
\end{equation}
where ${\cal L}_0(\Omega)$ is the Cooper pair propagator at the lowest Landau level given
by Eq.~(\ref{L3}), ${\cal L}_1$ is the Cooper pair propagator at the first Landau level,
which near $H_{\rm p2}(0)$ is not singular and can be treated as a constant, and $\gamma_{01}(\Omega,\omega)$
is a matrix element between the first and the lowest Landau level of the current
vertex]. The current vertex operator is defined by $\hat{{\bm \gamma}} (\Omega,
\omega)= {\bm \gamma} ({\bf q} \to {\bm \pi};
\Omega, \omega)$, where the latter quantity is given by the
three-Green's function block:
\begin{eqnarray}
\label{cvert} {\bm \gamma} ({\bf q} ; \Omega, \omega) &=& T
\sum\limits_{\varepsilon} \int {d^2 p \over (2 \pi)^2} {\bf v}\,
{\cal G}_{\varepsilon}({\bf p}) {\cal G}_{\varepsilon-\omega}({\bf
p}) {\cal G}_{\Omega-\varepsilon}({\bf q} - {\bf p}).
\end{eqnarray}
The matrix element $\gamma_{01}(\Omega,\omega)$  was calculated in
Ref.~\onlinecite{GaL} and reads
\begin{eqnarray}
\gamma_{01} (\Omega,\omega) = -{\nu r_H \over \sqrt{2}} \, {1
\over 1 + |\omega| \tau}   \left[ 1 - {\sqrt{\pi} \over 2} {r_H
\over v_F} \left( |\Omega| + |\Omega-\omega| + |\omega| \right)
\right] + \gamma_{\rm osc}, \label{vot}
\end{eqnarray}
where $r_H = 1/\sqrt{2eH}$ is the magnetic length for Cooper
pairs. Note that in Eq.~(\ref{vot}), we do not write explicitly the
oscillation contribution, $\gamma_{\rm osc}$. The oscillations coming from the
vertices are expected to have the usual Fermi liquid form, because
the two graphs corresponding to the two vertices when ``glued
together'' essentially reproduce the Drude conductivity diagram
and therefore their contribution can estimated modulo a
numerical coefficient in the overall amplitude.
Using the technique developed in Ref.~[\onlinecite{GaL}], we find the
following main result for the fluctuation conductivity, which
includes the leading oscillation terms as well (here, we address
only the zero-temperature contribution):
\begin{equation}
\label{DFosc} \delta \sigma = {e^2 \over  \pi^2} \left[ 1 + \eta
{\rho_{\rm osc} (H) \over \rho(0)} \right] \ln\left\{ {1 \over {\left[H
- \overline{H}_{\rm p2}(0)\right] / H} - C_{\rm osc}/\nu} \right\},
\end{equation}
with $C_{\rm osc}$ given explicitly by Eq.~(\ref{Cosc})\, $\eta\sim
1$ is a positive numerical coefficient, and $\rho_{\rm osc} (H)
/\rho (0)$ is the ratio of the Fermi liquid oscillation term and
the Drude resistivity in Eq.~(\ref{LK}). The latter does not
involve any additional factors in the temperature and
Dingle-temperature dependence as opposed to the Cooperon term,
which determines oscillations in the transition point
itself.~\cite{Tezan} We observe that Eq.~(\ref{DFosc}) describes a
decrease in the amplitude of the SdH oscillations upon approaching
the superconducting pairing instability.

Our result in Eq.~(\ref{DFosc}) may have an
interesting physical interpretation: The Cooper pair propagator
${\cal L}_n(\omega) = 1/\lambda_{\rm eff} \left\langle
\Delta^*(\omega) \Delta(\omega) \right\rangle$ corresponds to the
density of Cooper pairs at the $n$-th Landau level with the energy
$\omega$ (at least at $T=0$, the Matsubara frequency can be
converted into a real frequency via Feynman   rotation). Near
$H_{\rm p2}(0)$, only the Cooper pairs at the lowest Landau level
play a significant role and the total density of such pairs is
given by the integral over frequency [{\em cf.}, Eq.~(\ref{Qcl})]:
$N_{\rm  cp} \sim \int d\omega /(2 \pi) {\cal L}_0(\omega) \propto
-\ln\left\{ {H /[ H - H_{\rm p2}(0)]} \right\}$. Hence, the Cooper
pair density scales as a logarithm in the proximity to the
magnetic-field-tuned quantum transition and each electron within a
fluctuating pair produces an oscillation term. If temperature and
Dingle suppression are small, the functional dependence of the
oscillatory part of the fluctuation conductivity is dominated by
the Cooperon, $C_{\rm osc}$. The corresponding oscillation term
should also survive below the pairing transition in the gapless
superconductivity region. We note here that if such a term is
detected in experiment, then the application of the usual
Lifshitz-Kosevich formula to determine the effective carrier mass
from the temperature dependence of the amplitude of the
oscillations will {\em overestimate the effective  mass by the
factor of three}. In fact, there have been conflicting
experimental reports about the value of the effective mass for
carriers in the electron pocket and the above circumstance may be
relevant to this discrepancy. We expect that the temperature
dependence of the oscillation amplitude may exhibit a crossover
from $3 X(T) e^{-3\pi / (\omega_c \tau)} /\sinh \left[ 3 X(T)
\right]$ at relatively low fields (when most electrons are paired,
$H \lesssim H_{\rm p2}$)  to $X(T) e^{-\pi /(\omega_c \tau)} /
\sinh \left[ X(T) \right]$ at high fields (when most electrons are
unpaired, $H \gg H_{\rm p2}$).

\section{Two-Fluid Model and Transport Properties of the Pseudogap Phase}
\label{sec:propose}

Our main statement is that the pseudogap phase arises from the
Fermi surface reconstruction induced by antiferromagnetic
fluctuations into a holon Fermi surface in the nodal region
and an electron pocket in the anti-nodal region. The latter
electron pocket remains strongly paired  (but uncondensed) up to
very large energy scales, which may explain why
the underlying single-electron excitations had escaped discovery until
recently. The facts that the small electronic Fermi surface has
been ``visualized'' in the quantum oscillation experiments only in
a very narrow doping range and in just one class of materials are
most likely related to the purity of the samples. In addition, the
oscillations are observed in the vicinity of the ``magic'' doping
fraction, $p=1/8$, where it is known that stripe and other
competing orders are enhanced, which therefore (according to
the arguments of Sec.~\ref{sec:model}) should suppress the energy scales for
electron pairing and hence reduce the upper critical field at
which the oscillations are detectable. The latter argument is
consistent with the observation of a dip in the superconducting
dome in this doping region (we reiterate that the actual
superconducting transition temperature, $T_c$, corresponds to
Cooper pair condensation, which is different from the electron
pairing temperature, $T_p$. However, the values of two temperature
scales are expected to correlate strongly). It is therefore likely
that even though the electron pocket has not so far been directly
observed in quantum oscillation experiments and photoemission
measurements in other doping regimes and other materials, it does
exist throughout the phase diagram of the underdoped cuprates. Its
elusive nature can be explained by the exponential suppression of
the oscillations by the Dingle factor or/and much larger pairing
energy scales away from the magic doping fraction.

Because the paired electron pocket is argued to be central to the
phase diagram of the underdoped cuprates, it is important to
discuss whether other unusual properties of the pseudogap phase
(most importantly, its highly unusual transport properties)  are
consistent with the proposed mean-field state. In this section we
argue that indeed many such anomalous thermal and electric
transport experiments of the pseudogap phase can be qualitatively
understood within our picture. In particular, as proposed in the early
work of Geshkenbein {\em et al.},~\cite{geshkenbein} a change of sign in
transverse thermoelectric response can appear naturally within the
corresponding two-fluid model of a $-2e$ Cooper pair
Bose-liquid and the $+e$ hole Fermi liquid.
Here, we discuss a complementary formulation of this model by `dualizing'
the bosons into vortices, while retaining the gapless fermionic excitations of the
hole Fermi surface. Note that monopole Berry phases in $\mathcal{L}_A$ (see Appendix~\ref{app:gauge})
are likely to play an important role in the vortex action,~\cite{bbbss2,annals}
particularly in the structure of any charge order instabilities, but we will neglect this
complication here.

Among the most unusual experiments  in the pseudogap phase are the
Hall and Nernst measurements. Here we qualitatively discuss these two effects
within our two-fluid model of paired electrons and unpaired holon
excitations. For completeness, we discuss in Appendix~\ref{app:Tr} the
Hall conductivity of a generic anisotropic
Fermi liquid. There are also corresponding expressions for the Nernst coefficient,
as reviewed in Ref.~\onlinecite{hackl}.
In Appendices~\ref{app:Dual} and \ref{app:Boltz}, we review the dual mapping procedure and 
formally derive the conductivity composition rule of a vortex-holon two-fluid model:
\begin{equation}
\label{svf} \hat{\sigma} = {(2e)^2 \over h} \hat{\sigma}^{-1}_{\rm
V} +
\hat{\sigma}_{\rm f},
\end{equation}
where $\hat{\sigma}_{\rm V}$ is the dimensionless vortex
conductivity matrix and $\hat{\sigma}_{\rm f}$ is the electrical
conductivity matrix of negatively charged holons. Eqs.~(\ref{svf})
has the simple physical interpretation: In the pseudogap phase,
the electrons are paired but uncondensed, which means that
Cooper-pair conductivity, $\hat{\sigma}_{\rm CP}$, is finite and
is given by the vortex resistivity in the dual language.~\cite{MPAF} The total
conductivity is obtained simply by adding up the Cooper pair (or
vortex) and the holon Fermi-liquid contribution, which leads to
Eq.~(\ref{svf}). Using this equation, we get the following
expression for the Hall angle (see also,
Ref.~[\onlinecite{Fvort}], where this equation was first derived
in a different context):
\begin{equation}
\label{Hallangle} \tan{\theta_{\rm H}} = {\sigma_{\rm V}^{\perp} +
\sigma_{\rm f}^{\perp} {\rm Tr\,} \hat{\sigma_{\rm V}} \over
\sigma_{\rm V}^{||} + \sigma_{\rm f}^{||} {\rm Tr\,}
\hat{\sigma_{\rm V}}},
\end{equation}
where  ${\rm Tr\,} \hat{\sigma_{\rm V}} = \left( \sigma_{\rm
V}^{\perp} \right)^2 +  \left( \sigma_{\rm V}^{||}\right)^2$.  We
see that both the magnitude and the sign of the Hall response are
determined by the numerator of  Eq.~(\ref{Hallangle}), which
involves several physically distinct contributions that may and
most likely do have opposite signs. While the sign of holon Hall
conductivity is determined by that of the positive hole electric
charge, the sign of the vortex contribution is expected to be
opposite at least in the strong coupling regime. Indeed, assuming
a dilute density of electronic tightly bound pairs, we
expect them to behave like canonical bosons with charge ($-2e$), and
so contribute a negative Hall conductivity --- this is the basic
picture of Geshkenbein {\em et al.}~\cite{geshkenbein}

The issue of the vortex Hall angle becomes more settled in the
regime away from strong coupling. We emphasize here that by the ``vortex''
transport, here we are referring to the contribution of the electron pockets,
whose normal state transport is described in a dual model of vortices.
Within the framework of BCS
theory, vortex conductivity has been considered previously by
Dorsey~\cite{Dorsey} on the basis of the time-dependent
Ginzburg-Landau equation, which for the case of our electronic
pocket reads
\begin{equation}
\label{TDGL} \gamma \left(\partial_t - 2ie a_\tau \right) \Delta_g =
{\hbar^2 \over 4m^{\ast}} \left( {\bm \nabla} + 2ie{\bf a}
\right)^2 \Delta_g + \alpha \Delta_g - \beta |\Delta_g|^2 \Delta_g,
\end{equation}
where $\Delta_g (t,{\bf r})=  \lambda_{\rm eff} \left\langle g_+
g_- \right\rangle$ is the electronic Cooper-pair wave-function,
$\alpha$ and $\beta$ are the Landau parameters,  and $\gamma = \gamma_1 + i
\gamma_2$ is the order-parameter relaxation time. We note here
that the time-dependent Ginzburg-Landau theory formally has a very
narrow regime of applicability and is expected to be
quantitatively valid only in the gapless superconductivity region.
However, it should provide a useful insight about the vortex Hall
contribution in the crossover from strong to weak coupling.
According to Dorsey,~\cite{Dorsey} the vortex Hall angle is
related in a complicated way to the relaxation parameter,
$\gamma$, and the structure of the vortex core. But generally the
sign of the Hall angle (relative to that in the normal phase) is
determined by the sign of the following parameter,
$(-\gamma_2/\gamma_1)$. In the clean limit of unscreened
intervortex interactions (neutral superfluid) the situation
simplifies leading to $\gamma=-i$ (which basically ``restores''
the Gross-Pitaevskii-like equation for the Cooper pair fluctuations) and
giving the sign of the Hall angle identical to that in the normal
phase, which is consistent with the discussion above in terms of
the canonical bosons; the importance of this canonical boson contribution
to the Hall transport was pointed out by Geshkenbein {\em et al.}\cite{geshkenbein}.
In the weak-coupling and dirty limit, the
situation is different and the sign of the Hall angle is
determined by the Ginzburg-Landau time relaxation parameter given
by (here we present the result of the BCS model; see, {\em e.g.}, Aronov
{\em et al.\/}:~\cite{AHL})
\begin{equation}
\label{gamma} \gamma = {\pi  \over 8} - {i T_{\rm  p} \over 2}
{\partial \ln{T_{\rm p}} \over \partial \mu}.
\end{equation}
Note that the BCS weak-coupling limit is  not directly applicable
to the strongly-paired electron pocket. However, Eq.~(\ref{gamma})
above provides a tentative indication that upon approaching the
regime of weak Cooper pairing ({\em e.g.}, in the phase with enhanced
antiferromagnetism correlations), the sign of the vortex Hall
response may change. According to Eq.~(\ref{gamma}), it is
determined by the value of logarithmic derivative of the BCS
pairing temperature with respect to the chemical potential, which
is proportional to the derivative of the density of states at the
underlying Fermi surface (in our case, the Fermi surface of the
electron pocket); depending upon the shape of the Fermi surface,
this could have a negative sign.
Therefore, in the weak coupling limit, the signs of
the Hall contributions for the $h/(-2e)$-vortices and ($+e$)-holons
may become the same (which could, in principle, be relevant in the
vicinity of the ``magic doping level,'' where competing magnetic
orders are enhanced leading presumably to a suppression of
electron pairing, and where the change of sign in the Hall
response has been observed). We reiterate however that the
electron pocket is expected to remain strongly paired in the most,
possibly all, of the pseudogap phase and therefore generally the
vortex contribution is expected to retain the electron-like sign
and to compete with the holon contribution.

The overall sign in the experimentally observed Hall effect will
therefore be determined by the interplay of two physically
different terms, $\sigma_{\rm V}^{\perp}$ and $\sigma_{\rm
f}^{\perp}$, and can show reversals depending on the system
parameters. Again, in the proximity to the ``magic'' doping level,
one expects decreased disorder ({\em i.e.\/}, decreased pinning strength)
and suppression of the electron pairing as well, which should
significantly alter the vortex contribution. We note that
Taillefer {\em et al.\/}~\cite{louis} have reported a strong
correlation between the sign-reversal in the Hall effect and the
presence of quantum oscillations, which is qualitatively
consistent with the afore-mentioned scenario.

In the simplest application of the two-fluid model, the Nernst
effect will be determined by the contribution of the
$h/(-2e)$-vortices and the holons. However, in
reality the situation may be more complicated due to a strong
(possibly competing) effect of the superconducting
fluctuations~\cite{NernstG} (here we imply Aslamazov-Larkin
amplitude fluctuations), which are known to be large compared to
the Fermi-liquid terms even far above the pairing transition.
Furthermore, possible charge-ordering instabilities of the vortex liquid,~\cite{bbbss}
and the associated proximity to the insulating state at $p=1/8$ likely
also play a role.~\cite{markus}

\section{Conclusions}
\label{sec:Conc}

This paper has combined insights from recent experiments \cite{shen1,kanigel,letacon,shen2,kohsaka1,hudson,kohsaka2,vidya,sawatzky,doiron,louis,cooper,nigel,cyril,suchitra} and earlier theoretical developments
\cite{rkk1,geshkenbein,rkk2,rkk3} to present a fairly simple model for the underdoped cuprates with a firm microscopic and theoretical foundation. We start with a Fermi liquid state with long-range SDW order, containing electron
pockets near the ${\bf G}_a$ wavevectors, and hole pockets near the ${\bf K}_v$ wavevectors (see Fig.~\ref{fig:bz}).
Then we express the spin polarization of the electronic excitations near these pockets in terms of the {\em local\/}
polarization of the SDW order; this is the content of Eqs.~(\ref{F}) and (\ref{G}). The advantage of this procedure
is that it allows to easily extend key aspects of the physics of the small pockets into the phase without long-range
SDW order. In particular, it shows that the electronic excitations experience a long-range gauge force associated
with an emergent U(1) gauge field ${\bf A}$.

A key feature of our theory is that the primary pairing instability is associated
with electron-like pockets near the ${\bf G}_a$ wavevectors.
We showed that these pockets experience a strong attractive
pairing force from the transverse gauge fluctuations, and this leads naturally to an $s$-wave pairing instability.
However, after rotating back to the physical spin polarization direction via Eq.~(\ref{G}), the resulting paired
state was found to have a $d$-wave pairing signature for the physical electrons. Next we focused attention
on the Josephson couplings in Eq.~(\ref{Josephson})  between the electron and hole pockets: we found
that it induced a $p$-wave pairing of the holons, which was strongly frustrated by the gauge forces on the holons.
Again, after rotating back to the physical electrons using Eq.~(\ref{F}), this very weakly paired holon
state was found to have a $d$-wave pairing signature with nodal fermionic excitations.
The ``nodal-anti-nodal dichotomy'' is a natural consequence of this theory, with very different pairing physics near
the ${\bf G}_a$ and ${\bf K}_v$.

Section~\ref{sec:qosc} explored the nature of the SdH oscillations in the normal state induced by a strong
magnetic field at low temperatures. We computed the nature of the suppression of these oscillations
by gauge field and pairing fluctuations.

Section~\ref{sec:propose} explored aspects of transverse transport in the finite temperature pseudogap
phase. Here, we also discussed connections to the boson-fermion
model of Geshkenbein {\em et al.}~\cite{geshkenbein}.

We conclude by mentioning that to determine the
correctness of our proposed theory for the underdoped cuprates,
it would be essential to visualize the single-particle excitations in
the electron pocket in other experiments, apart from the existing quantum
oscillation measurements.  Such new experiments should involve external perturbations,
which {\em destroy superconducting pairing} in the electron pocket, without smearing out
its small Fermi surface or altering the underlying magnetic or topological order that yields
Fermi surface reconstruction.  Due to the former limitation, high temperature and/or
strong disorder are not appropriate for this purpose. Possible other means could be, {\em e.g.},
to study AC transport in a magnetic field, looking, in particular,
for cyclotron resonance effects coming from the single-electron excitations.
Since the quantum oscillations have been observed, the materials are
sufficiently clean to exhibit the cyclotron resonance phenomena as
well. Another promising avenue could be to experimentally investigate
{\em non-linear} transport, {\em e.g.}, non-linear $IV$-curves
in a magnetic field. In the vicinity of the upper critical field $H_{\rm p2}(T)$, the
critical (depairing) current is expected to be
relatively small and some manifestations of
single-electron physics would appear at {\em lower fields} as
compared to linear transport. Of particular interest would be also
to compare the behavior of non-linear $IV$-curves in the regions
with the opposite signs of the Hall conductivity. In our theory, the sign reversal
in Hall response occurs naturally due to the competing contributions from the
uncondensed  electronic Cooper pairs and holons. This competition combined with the
effects of depairing may lead a non-monotonic behavior and, possibly, sign reversals
in non-linear transport data.

\acknowledgements

We are very grateful to Ady Stern for a
discussion in which he pointed out the connection to the pairing
problem in double layers of $\nu=1/2$ quantum Hall systems.~\cite{bmn,ady} We
thank Seamus Davis, Dennis Drew, Antoine Georges, Rick Greene, Andreas Hackl, Lev Ioffe,
Yong-Baek~Kim, Markus M\"uller, John-Pierre Paglione, Suchitra Sebastian, \mbox{Z.-X.}~Shen, Tudor
Stanescu, and Louis Taillefer for useful discussions.
 VG acknowledges the NSF Physics Frontier Center at JQI.
 The research of SS was
supported by the NSF under grant DMR-0757145 and by the FQXi
foundation.

\appendix

\section{Complete Lagrangian}
\label{app:l}

Only two of the terms in the Lagrangian in Eq.~(\ref{L}), $\mathcal{L}_g$ and $\mathcal{L}_{fg}$, were displayed
in the body of the paper. This appendix will display the remaining terms,
along with a brief discussion
of their physical consequences.

\subsection{Spinons}
The Lagrangian for the $z_\alpha$ is \cite{rkk1}
\begin{equation}
\mathcal{L}_z = |( \partial_\mu - i A_\mu ) z_\alpha
|^2 + s |z_\alpha |^2 + \frac{u}{2} \left( |z_\alpha|^2 \right)^2,
\label{Lz}
\end{equation}
and the spinon ``mass'' term $s$ tunes a transition from the SDW
ordered state ($\langle z_\alpha \rangle \neq 0$) to a state with
spin rotation invariance preserved ($\langle z_\alpha \rangle =
0$). Note that in the SDW ordered
state, the spinon condensate induces a Higgs ``mass'' term, $|\langle z_\alpha \rangle |^2 A_\mu^2$, for the $A_\mu$.

\subsection{Holons}
These are associated with the
$F_{v\alpha}$ fermionic excitations near the ${\bf K}_v$
wave-vectors \cite{rkk2}
\begin{equation}
\mathcal{L}_f =  \sum_{q=\pm} \sum_{p=1,2} f^\dagger_{qp}  \biggl[
\partial_\tau - i q A_\tau  -i e a_\tau + \mu - E_{fg} - \frac{(\partial_{\overline{j}} - i
q A_{\overline{j}} - i e a_{\overline{j}} )^2}{2m_{p\overline{j}}} \biggr] f_{qp},
\label{Lf}
\end{equation}
where $\overline{j}$ extends over $\overline{x},\overline{y}$,
$m_{1\overline{x}}=m_{2\overline{y}}$ and
$m_{2\overline{x}}=m_{1\overline{y}}$ are the masses of the
elliptical holon pockets, and the $\overline{x}$ and
$\overline{y}$ directions are rotated by 45$^\circ$ from the
principle square axes. Eq.~(\ref{Lf}) also contains the energy
$E_{fg}$, which is the analog of the ``semiconductor'' band gap,
between the top of the hole (valence) band and the bottom of the
electron (conduction) band. We expect that the value of $E_{fg}$
is sensitive to the strength of the SDW order, decreasing as the
SDW order becomes larger.

\subsection{Spin-charge couplings}
These couple the
spinons $z_\alpha$ to the holons $f_{\pm p}$ and the doublons
$g_{\pm}$. The simplest allowed terms are  couplings between the
scalar densities, $|z_\alpha|^2$ and $f^\dagger_{qp} f_{\pm p}$,
$g_\pm^{\dagger} g_\pm$. However, there are also
``Shraimain-Siggia'' terms \cite{ss} which couple operators
carrying charges $\pm 2$ under $A_\mu$. Again, the most general
form of all these terms can be deduced from the PSG, as has been
described in previous works. For the holons, the spin-charge
couplings take the form
\begin{equation}
\mathcal{L}_{zf} = \lambda_{zf} |z_\alpha|^2 \sum_{qp} f^\dagger_{qp} f_{qp} +   i \widetilde{\lambda}_{zf} \varepsilon^{\alpha\beta}
\left\{ f^\dagger_{+1} f_{-1} z_\alpha
\partial_{\overline{x}} z_\beta  + f^\dagger_{+2} f_{-2} z_\alpha
\partial_{\overline{y}}  z_\beta  \right\} + \mbox{H.c.}.
\label{eqSS}
\end{equation}
The second term is the Shraiman-Siggia term; in the SDW state, this term favors incommensurate spiral
spin correlations. In the non-magnetic state with $\langle z_\alpha \rangle = 0$, this term moves the electron spectral weight
away from the commensurate ${\bf K}_v$ points, to be centered on a ``Fermi arc'', as has been described in previous
work.~\cite{rkk1} Finally, integrating out the $z_\alpha$ also leads to an attractive pairing term \cite{qi2} for the $f_{\pm p}$.
For the $g_\pm$, the spin-charge couplings are
\begin{eqnarray}
\mathcal{L}_{zg}&=& \lambda_{zg} |z_\alpha|^2 \sum_q g^\dagger_q g_q \nonumber \\ &+&
\widetilde{\lambda}_{zg} \biggl\{ ~\varepsilon^{\alpha\beta}\left[g_+^\dagger  \left( D_x g_- \right) z_\alpha \left(D^-_x z_\beta \right)
-g_+^\dagger \left( D_y g_- \right) z_\alpha \left(D_y z_\beta \right) \right] \\
&~&~~~~~~~+ \varepsilon_{\alpha\beta}\left[g_-^\dagger \left(D_y g_+ \right) z^{\alpha*} \left(D_y z^{\beta *} \right)
- g_-^\dagger \left(D_x g_+ \right) z^{\alpha*} \left(D_x z^{\beta *} \right)\right] \biggr\} + {\rm H.c.}.\nonumber
\label{lc2}
\end{eqnarray}
Now the Shraiman-Siggia term has 2 spatial gradients and
does not induce spiral correlations.

\subsection{Gauge field}
\label{app:gauge}
These are induced by
integrating out the matter fields, and can take a different form
depending upon the nature of the matter excitations. When Fermi
surfaces for $g_\pm$ are present, the gauge field dynamics is
overdamped, as discussed in the body of the paper. However, in all
phases,  terms of the form $\mathcal{L}_A \sim (\partial_\mu A_\nu
- \partial_\nu A_\mu)^2$ are always allowed, obtained by
integrating high energy degrees of the freedom. For the crossover
to the confining state, we also need to consider topologically
non-trivial configurations of the $A_\mu$ corresponding to
monopole tunnelling events. These have been ignored in the present
paper because the monopoles are suppressed by the holon Fermi
surfaces, but their effects have been discussed earlier
\cite{rkk1,rkk2,rkk3} in some detail. The monopoles come with
Berry phases, and these are crucial in determining the nature of
translational symmetry breaking in the confining phases.

\section{Hall Effect in an Anisotropic Fermi Liquid} \label{app:Tr}

In this appendix, we summarize the properties of transverse
thermoelectric linear response in a two-dimensional anisotropic
Fermi liquid and present a general expression for the Hall coefficient.
The results presented in this
appendix are used in Sec.~\ref{sec:propose}, where the transverse
thermoelectric response in the pseudogap phase is discussed.

Consider a two-dimensional Fermi liquid with the anisotropic
dispersion $E = E({\bf p})$. We also introduce the following
standard notation: $\xi(p,\phi_{\bf p}) = E(p,\phi_{\bf p}) -
E_{\rm F}$, where ${\bf p} = (p,\phi_{\bf p})$ is simply the
momentum in polar coordinates and $E_{\rm F}$ is the Fermi energy.
This indirectly determines the value of the particle momentum as
the function of energy and angle: $p = p(\xi,\phi_{\bf p})$. One
can also define the particle velocity as a function of the energy
and the angle: ${\partial E / \partial p_{\alpha}} =
v_{\alpha}(\xi,\phi_{\bf p})$. We now introduce the following
identity to treat the integrals over momentum [below, $F({\bf p})$
is arbitrary function]
\begin{equation}
\label{IF} I_{\rm F} = g_{\rm s} \int{d^2p \over (2\pi)^2} F({\bf
p}) = \int d\xi \left\langle \nu(\xi,\phi_{\bf p}) F(\xi,\phi_{\bf
p})  \right\rangle,
\end{equation}
where $g_s$ is a degeneracy due to an internal degree of freedom
({\em e.g.}, spin in a usual electron liquid or a sublattice index in
our theory), $\langle \ldots \rangle = \int_0^{2\pi} (d\phi /2\pi)
\ldots$ is the average over the directions in the Brillouin zone
(which reduces to the average over the Fermi surface if $\xi=0$),
and we introduced the density of states:
\begin{equation}
\label{nu}  \nu(\xi,\phi_{\bf p}) = {g_{\rm s} \over 4\pi}
{\partial \over \partial \xi} p(\xi,\phi_{\bf p}).
\end{equation}
Let us also introduce the following notation for the integral that
often appears in deriving the finite temperature transport
properties of a Fermi liquid:
\begin{equation}
\left[ f(\varepsilon) \right]_T = \int\limits_{-\infty}^{\infty}
{d \varepsilon \over 4T} {f(\varepsilon) \over
\cosh^2\left\{\varepsilon /( 2T)\right\}}. \label{avT}
\end{equation}
At zero temperature, it simply gives $\left[ f(\varepsilon)
\right]_{T\to 0} = f(0)$. The notations defined by
Eqs.~(\ref{IF}), (\ref{nu}), and (\ref{avT}) will be used below to
express the general thermoelectric response coefficients in a
compact and intuitive form, which would allow a simple physical
interpretation.

The general theory of the Hall coefficient within the Green's
function formalism has been considered by many authors and we
refer the reader to the corresponding literature (see, {\em e.g.},
Altshuler and Aronov~[\onlinecite{AAr}] and
Livanov~[\onlinecite{Livanov}]). In the general case of an
anisotropic Fermi liquid with an angle-dependent scattering time,
$\tau_{\phi_{\bf p}} (\varepsilon)$, one can obtain the following
expression for the Hall conductivity in the limit of a weak
magnetic field ({\em i.e.\/}, if $\langle \omega_c \tau \rangle \ll 1$):
\begin{equation}
\label{Hall} \sigma^{\perp}_{\alpha \beta} = e^3 H \left[
\left\langle \tau_{\phi_{\bf p}}^2(\varepsilon) {\partial \over
\partial \varepsilon} \left\{ v_{\alpha}^2(\varepsilon,\phi_{\bf
p}) v_{\beta}^2(\varepsilon,\phi_{\bf p})
\nu(\varepsilon,\phi_{\bf p}) \right\} \right\rangle \right]_T.
\end{equation}
In the zero temperature limit, $T/E_{\rm F} \to 0$, and assuming
an angle-independent scattering time, $\tau$, we get the
simplified equation for the Hall response ($\alpha \ne \beta$):
\begin{equation}
\label{Hall1} \sigma^{\perp}_{\alpha \beta} = e^3 \tau^2 H
\left\langle  {\partial \over \partial \varepsilon} \left\{
v_{\alpha}^2 v_{\beta}^2 \nu \right\} \right\rangle_{\rm FS}.
\end{equation}
The corresponding expression for the longitudinal Drude
conductivity is
\begin{equation}
\label{Drude} \sigma^{||}_{\alpha \alpha} = e^2  \tau
\left\langle v_{\alpha}^2 \nu  \right\rangle_{\rm FS}.
\end{equation}
In the isotropic limit of a circular Fermi surface, $v^2 = 2E/m$,
and Eq.~(\ref{Hall1}) reproduces the familiar expression for the
Hall conductivity $\sigma^{\perp} = ({\omega_c \tau})
\sigma^{||}$, with $\sigma^{||} = (v_{\rm F}^2 /2) \nu e^2 \tau$ being the
longitudinal Drude conductivity of Eq.~(\ref{Drude}).  We
emphasize here that according to Eq.~(\ref{Hall1}), the Hall
coefficient does depend on the derivative of the density of
states, but the latter effects are not necessarily dominant. {\em E.g.},
the Hall conductivity is finite even if the density-of-states is a
constant. The sign of the Hall effect can change within a
single-band Fermi liquid picture only if the density-of-states
depends on the energy stronger than $v_x^2 v_y^2$ in the
corresponding directions, which requires a very anisotropic Fermi
surface.

\section{Duality Transformation}
%
\label{app:Dual}

The mean-field Lagrangian of this two-fluid model can be written
as follows:
\begin{equation}
\label{L2f} {\cal L}[\Delta_g,f] = {\cal L}_{\rm GL}[\Delta_g] +
{\cal L}_{\rm h}[f] + {\cal L}_{\rm J} [\Delta_g,f],
\end{equation}
where $\Delta_g = \lambda_{\rm eff} \left\langle g_+ g_-
\right\rangle$ is the pairing order parameter describing electron
Cooper pairs and ${\cal L}_{\rm GL}$ is the corresponding
Ginzburg-Landau Lagrangian
\begin{equation}
\label{LDg} {\cal L}_{\rm GL}[\Delta_g] =   \Delta_g^* \left[
{\left( -i{\bm \nabla} - 2e{\bf a} \right)^2 \over 4m}
+\left(\partial_\tau - 2ie a_\tau \right) \right] \Delta_g +
\alpha(T) \left| \Delta_g \right|^2 - \beta \left| \Delta_g
\right|^4,
\end{equation}
where $a = \left(a_\tau, {\bm a}\right)$ is the physical electromagnetic field (we assumed
that all effects of the gauge filed, $A$, have already been
incorporated into the effective parameters) and $\alpha$ and
$\beta$ are Ginzburg-Landau parameters, which determine the
modulus of the order parameter, $|\Delta_g| \equiv \Delta_0 =
\sqrt{\alpha/(2\beta)}$, which we assume fixed. However, the
phase, $\phi$, of the order parameter, $\Delta_g = \Delta_0
\exp(i\phi)$, is allowed to fluctuate. This leads to the XY-model
for the electronic Cooper pairs and a Kosterlitz-Thouless
transition below the BCS pairing temperature, $T_{\rm p0}$.
The second term in Eq.~(\ref{L2f}) is the free fermion Lagrangian
describing the motion of holes
\begin{equation}
\label{Lh} {\cal L}_{\rm h}[f] = f^* \left[ {\left( -i{\bm \nabla}
+ e{\bf a} \right)^2 \over 2 m_h} + \left(\partial_\tau + ie a_\tau
\right) \right] f,
\end{equation}
where the term $(+e{\bf a})$ describes coupling of the holes to
the electromagnetic field. Finally, the last term in
Eq.~(\ref{L2f}), ${\cal L}_{\rm J} [\Delta_g,f]$, is given by
Eq.~(\ref{Josephson}) and describes internal tunnelling between
the electrons and the holes.

The structure of the tunnelling term is given by $e^{i\phi} ff$,
and it is very similar to that in the usual BCS mean-field model
of a gapless superconductor (see, {\em e.g.},
Ref.~[\onlinecite{Fvort}]). One can therefore just follow the
steps used in Refs.~[\onlinecite{Fvort,rotons}] to derive the
so-called $U(1)$ formulation~\cite{rotons} of the vortex-fermion
mixture in this model, which describes the motion of vortices
statistically coupled to gapless fermions: The statistical
interaction is that they ``see each other'' as sources of a
$(-\pi)$-flux and therefore induce electromotive force on each
other when moving. The first purely technical step is to introduce
a new operator, $h_{\bf r} = e^{i\phi_{\bf r}/2} f_{\bf r}$, which
simplifies the Josephson term. The next step is to perform a
duality transformation with respect to the bosonic Cooper-pair
field, $e^{i\phi_{\bf r}}$. The resulting action describing a
two-fluid vortex-holon liquid is
\begin{eqnarray}
\label{Lvh} {\cal L}[\Psi_V,h] =  && \Psi_V^* \left[ {\left(
-i{\bm \nabla} - {\bf a}^{\rm dual} + {\bm \alpha} \right)^2 \over
2M_V} + \left(\partial_\tau + i{a}^{\rm dual}_\tau + i\alpha_\tau\right)
\right] \Psi_V\\
&& +h^* \left[ {\left( -i{\bm \nabla} - {\bm \beta} \right)^2
\over 2 m_h} + \left(\partial_\tau  -i\beta_\tau \right) \right] h
+ {\cal L}_{\rm gauge},
\end{eqnarray}
where $M_{\rm V}$ is a vortex mass, $a^{\rm dual}$ is the gauge
field, which describes the Cooper pair density fluctuations [${\bm
\nabla} \times {\bf a}^{\rm dual} = n_{\rm CP} ({\bf r})$], and the only
purpose of the fields ${\bf \alpha}$ and ${\bf \beta}$ is to
mediate the long-range statistical interaction between the
vortices and the fermions. These fields are ``attached'' to the
vortices via ${\bm \nabla} \times {\bm \beta} = i\pi \Psi_{\rm
V}^\dagger \Psi_{\rm V}$ and to the fermions via ${\bm \nabla}
\times {\bm \alpha} = i\pi h^\dagger h$ This statistical
interaction is described by the mutual Chern-Simons term, which is
the first term in the following gauge part of the action
\begin{equation}
\label{Lgauge} {\cal L}_{\rm gauge} = -{i \over \pi} \left(
\epsilon_{\mu\nu\lambda} \alpha_\mu \partial_\nu \beta_\lambda
\right)  + {1 \over 2C} \left(\epsilon_{\mu\nu\lambda}
\partial_\nu a^{\rm dual}_\lambda  - \delta_{\mu 0} 2 \pi n_{0}
\right)^2  + {i \over 2\pi} J^{\rm cp}_\mu
\epsilon_{\mu\nu\lambda} \partial_\nu a^{\rm dual}_\lambda .
\end{equation}
The second term in Eq.~(\ref{Lgauge}) describes dual gauge-field
fluctuations, which physically correspond to plasmons. In a
charged system, the plasmons have a gap due to the long-range
Coulomb forces and as such the fluctuations of this gauge field
are expected to be less pronounced than in a neutral Bose-system.
Finally, the last term describes coupling to a physical electric
Cooper pair current.

The theory of Eqs.~(\ref{Lvh}) and (\ref{Lgauge}) summarizes the
following essential features of the two-fluid vortex-holon
mixture: The vortices and the Cooper pairs ``see each other'' as
$(+2\pi)$-fluxes and induce EMFs (transverse Magnus forces) on
each other when moving. Likewise, the vortices and the gapless
fermions (holons) ``see each other'' as $(-\pi)$-fluxes and induce
EMFs as well. We will be using this picture in derivation of the
semiclassical transport equations below.

A more settled issue  is the question of the total effective
magnetic field (``dual field'') seen by a vortex. According, to
Eqs.~(\ref{Lvh}) and (\ref{Lgauge}), it is $B^{\rm (dual)} = {\bm
\nabla} \times \left[ {\bf a}^{\rm (dual)} - {\bm \alpha} \right]
= 2\pi \left( n_{0} - {1 \over 2} n_{\rm h} \right)$. We reiterate
that the Josephson term~(\ref{Josephson}) in the action violates
the individual conservation laws for the $g$- and $f$-particles.
This means in particular that the density of the latter, $n_h$,
may vary depending on the phase.  Another related non-trivial
question is about the vortex statistics (with respect to each
other). The direct duality transformation gives bosonic vortices,
however other statistics are in principle possible. These are very
interesting questions, which however are beyond the scope of the
present study. Below, we treat the vortices semiclassically to
develop a phenomenological theory of transport and derive the
corresponding transport coefficients.

\section{Conductivity Composition Rule in the Two-Fluid Vortex-Holon Model}
\label{app:Boltz}

The derivation of the semiclassical theory of transport in the
two-fluid model is essentially identical to that of Ref.~[\onlinecite{Fvort}]
and is based on the following equations,
which describe the electromotive forces between the vortices,
fermions, and Cooper pairs in the presence of currents and thermal
gradients:
\begin{equation}
\label{qc}
\begin{array}{c}
{\bf j}_{\rm v}=-\hat{\sigma}_{\rm v} \hat{\epsilon} \left(
{\bf j}_{\rm h} + {\bf j}_{\rm CP} \right) - \hat{\lambda}_{\rm v} {\bm \nabla} T;\\
{\bf j}_{\rm h}= -\hat{\sigma}_{\rm h} \hat{\epsilon} {\bf j}_{\rm
v}- \hat{\lambda}_{\rm h} {\bm \nabla} T,
\end{array}
\end{equation}
where $\hat{\epsilon}$ is the antisymmetric tensor in two
dimensions, ${\bf j}_{\rm v}$, ${\bf j}_{\rm h}$, and ${\bf
j}_{\rm cp}$ are the vortex, holon, and Cooper pair density
currents respectively and $\hat{\sigma}_{\rm v/h}$ and
$\hat{\lambda}_{\rm v}$ are the vortex/holon dimensionless
conductance and thermal conductivity matrices respectively. The
latter two matrices generally have the form
\begin{equation}
\begin{array}{cc} \sigma=\left(\begin{array}{cc} \sigma_{||} & \sigma_{\perp} \\
-\sigma_{\perp} & \sigma_{||} \end{array} \right) & \mbox{ and }
\lambda=\left(\begin{array}{cc} \lambda_{||} & \lambda_{\perp} \\
-\lambda_{\perp} & \lambda_{||} \end{array} \right).
\label{conducts}
\end{array}
\end{equation}
The quantities of interest are the total electrical
and thermal conductivity matrices for the system. {\em E.g.}, in the
absence of thermal gradients, the conductivity tensor is defined
by ${\bf j}_{\rm CP} = \hat{\epsilon} {\bf E}$, while the actual
electric field is determined by ${\bf E} = {1 \over 2e}
\hat{\epsilon} {\bf j}_{\rm v}$. Then, Eqs.~(\ref{qc}) can be
easily resolved and give the following expression for the total
electrical conductivity matrix:
\begin{equation}
\label{sigma} \hat{\sigma} = (2e)^2 \left[ \hat{\sigma}^{-1}_{\rm
V} + {1 \over 4} \hat{\sigma}_{\rm h} \right].
\end{equation}
Despite the rather complicated set of arguments and transformation
that have led to this result, the physics of Eq.~(\ref{sigma}) is
very simple: The total electrical conductivity in the uncondensed
liquid phase is given by the sum of Cooper pair and hole
conductivities. The former can be related to the vortex transport
properties and due to duality is simply given by the vortex
resistivity. One can also derive a complete phenomenological
expression for the Peltier tensor in the two-fluid model defined
via ${\bf E} = \hat{\lambda} {\bm \nabla} T$. The exact expression
for the Peltier tensor is rather involved, but assuming that the
hole contribution to the Nernst effect is negligible, one can use
the results of Ref.~[\onlinecite{Fvort}].


\begin{thebibliography}{}
\bibitem{ginzburg} Note that our estimate of the Ginzburg parameter
is for the small pockets of electrons, near the anti-nodal points.
In contrast, Ref.~\onlinecite{geshkenbein} assumed these electrons had already paired into bosons.
They estimated a very small Ginzburg
parameter for the fermions near the Fermi arcs along the Brillouin zone diagonals;
the pairing of these fermions is also relatively weak in our theory.
\end{thebibliography}
\end{document}